\documentclass[]{nature}
\makeatletter\if@twocolumn\PassOptionsToPackage{switch}{lineno}\else\fi\makeatother

\newcommand{\widesim}[2][1.5]{
  \mathrel{\overset{#2}{\scalebox{#1}[1]{$\sim$}}}
}

\newcommand{\toover}[2][1.5]{
  \mathrel{\overset{#2}{\scalebox{#1}[1]{$\to$}}}
}

\usepackage{tabulary,graphicx,amsmath,amsfonts,amssymb}
\usepackage[utf8x]{inputenc}
\usepackage{amsfonts}
\usepackage{amssymb}
\usepackage{amsmath}
\usepackage{soul}
\usepackage{MnSymbol}
\usepackage{graphicx}
\usepackage{color}
\usepackage{lineno}

\makeatletter
\renewenvironment{figure}
               {\@float{figure}}
               {\end@float}
\renewenvironment{figure*}
               {\@dblfloat{figure}}
               {\end@dblfloat}

\renewenvironment{table*}
               {\@dblfloat{table}}
               {\end@dblfloat}

\makeatother

\usepackage{url,multirow,morefloats,floatflt,cancel,tfrupee}
\makeatletter

\AtBeginDocument{\@ifpackageloaded{textcomp}{}{\usepackage{textcomp}}}
\makeatother
\usepackage{colortbl}
\usepackage{xcolor}
\usepackage{pifont}
\usepackage[nointegrals]{wasysym}
\makeatletter

\def\mcWidth#1{\csname TY@F#1\endcsname+\tabcolsep}

\def\cAlignHack{\rightskip\@flushglue\leftskip\@flushglue\parindent\z@\parfillskip\z@skip}
\def\rAlignHack{\rightskip\z@skip\leftskip\@flushglue \parindent\z@\parfillskip\z@skip}

\@ifundefined{etal}{}{}

\usepackage{ifxetex}
\ifxetex\else\if@twocolumn\@ifpackageloaded{stfloats}{}{\usepackage{dblfloatfix}}\fi\fi

\AtBeginDocument{
\expandafter\ifx\csname eqalign\endcsname\relax
\def\eqalign#1{\null\vcenter{\def\\{\cr}\openup\jot\m@th
  \ialign{\strut$\displaystyle{##}$\hfil&$\displaystyle{{}##}$\hfil
      \crcr#1\crcr}}\,}
\fi
}

\AtBeginDocument{%
  \@ifpackageloaded{endfloat}%
   {\renewcommand\efloat@iwrite[1]{\immediate\expandafter\protected@write\csname efloat@post#1\endcsname{}}}{\newif\ifefloat@tables}%
}%

\def\BreakURLText#1{\@tfor\brk@tempa:=#1\do{\brk@tempa\hskip0pt}}
\let\lt=<
\let\gt=>
\def\processVert{\ifmmode|\else\textbar\fi}

\@ifundefined{subparagraph}{
\def\subparagraph{\@startsection{paragraph}{5}{2\parindent}{0ex plus 0.1ex minus 0.1ex}%
{0ex}{\normalfont\small\itshape}}%
}{}

\newcommand\role[1]{\unskip}
\newcommand\aucollab[1]{\unskip}
  
\@ifundefined{tsGraphicsScaleX}{\gdef\tsGraphicsScaleX{1}}{}
\@ifundefined{tsGraphicsScaleY}{\gdef\tsGraphicsScaleY{.9}}{}
\def\checkGraphicsWidth{\ifdim\Gin@nat@width>\linewidth
	\tsGraphicsScaleX\linewidth\else\Gin@nat@width\fi}

\def\checkGraphicsHeight{\ifdim\Gin@nat@height>.9\textheight
	\tsGraphicsScaleY\textheight\else\Gin@nat@height\fi}

\def\fixFloatSize#1{}
\let\ts@includegraphics\includegraphics

\def\inlinegraphic[#1]#2{{\edef\@tempa{#1}\edef\baseline@shift{\ifx\@tempa\@empty0\else#1\fi}\edef\tempZ{\the\numexpr(\numexpr(\baseline@shift*\f@size/100))}\protect\raisebox{\tempZ pt}{\ts@includegraphics{#2}}}}

\AtBeginDocument{\def\includegraphics{\@ifnextchar[{\ts@includegraphics}{\ts@includegraphics[width=\checkGraphicsWidth,height=\checkGraphicsHeight,keepaspectratio]}}}

\DeclareMathAlphabet{\mathpzc}{OT1}{pzc}{m}{it}

\def\URL#1#2{\@ifundefined{href}{#2}{\href{#1}{#2}}}

\def\UrlOrds{\do\*\do\-\do\~\do\'\do\"\do\-}%
\g@addto@macro{\UrlBreaks}{\UrlOrds}

\edef\fntEncoding{\f@encoding}

\makeatother

\newif\ifmultipleabstract\multipleabstractfalse%
\usepackage[nomarkers,tablesfirst,nolists]{endfloat}

\def\fixFloatSize#1{}

\usepackage{float}

\begin{document}

\title{Probing the Unruh effect with an accelerated extended system}

\author{Cesar A. Uliana Lima$^{1}$\hspace{.2pc},
              Frederico Brito$^{1}$\hspace{.2pc},
              Jos\'{e} A. Hoyos$^{1}$\hspace{.2pc} \&
              Daniel A. Turolla Vanzella$^{1}$\thanks{Corresponding author.}\hspace{.4pc}\thanks{E-mail: vanzella@ifsc.usp.br}~\hspace{.2pc}
    }

\maketitle 

\begin{affiliations}
  \item 
    Instituto de F\'{\i}sica de S{\~{a}}o Carlos, Universidade de S{\~{a}}o Paulo\unskip, Caixa Postal 369\unskip, S{\~{a}}o Carlos\unskip, 13560-970\unskip, S{\~{a}}o Paulo\unskip, Brazil
\end{affiliations}

\begin{abstract}
It has been proved in the context of  quantum fields
in Minkowski spacetime that the vacuum state
is a  thermal  state according to uniformly accelerated observers --- a  seminal result known as 
the Unruh effect. 
Recent claims, however, have challenged the validity of this result for
extended systems, thus casting doubts on its physical reality. Here, we study the dynamics of an
 extended system,
uniformly accelerated
in the vacuum.
We show that its reduced density matrix evolves to a 
Gibbs thermal state with local temperature given by the Unruh 
temperature  $T_{\rm U} = \hbar a/(2\pi c k_{\rm B})$, 
where $a$ is the system's spatial-dependent
proper acceleration
— $c$ is the speed of light and $k_{\rm B}$ and $\hbar$ are the Boltzmann's and the reduced Planck's constants, respectively.
This proves that 
the vacuum state does induce thermalization 
of an accelerated extended system --- which is all one can expect of a legitimate thermal reservoir.
\end{abstract}\def\keywordstitle{Keywords}

    \textbf{\keywordstitle:} Unruh effect, Accelerated spins, Thermalization

\section{Introduction}

Soon after S.\ Hawking published his seminal result on particle creation due to black hole formation, 
leading to the
phenomenon of black hole evaporation\cite{Hawk}, W.~Unruh clarified the relative character of the  
 particle concept in the context of quantum field theory in flat spacetime.
More specifically, he showed that the vacuum state --- which 
represents absence of particles according to inertial observers --- corresponds to a thermal bath with temperature
$T_{\rm U} = \hbar a/(2 \pi c k_{\rm B})$ for uniformly accelerated observers\cite{Unruh}, 
where $a$ is the observers'   proper  acceleration
($\hbar$ is the reduced Planck's constant, $c$ is the speed of light, and $k_{\rm B}$ is the Boltzmann's constant).
This result has become known as the Unruh effect (see Ref.\cite{CHM} for a comprehensive
review). Although some deep connections may be established 
between Hawking's result
and the Unruh effect --- and, in fact, the former served as motivation for Unruh's analysis ---, the latter
is not as well known as the former. This is somewhat unfortunate because some conceptual issues
raised by black hole evaporation can be better understood through the lens of Unruh's result  
--- e.g.,  that the possibility of Hawking radiation being in a mixed state does  not violate any  quantum principle
(therefore, no information-loss ``paradox'' is present\cite{UW}). But even
among those who are acquainted with the Unruh effect, not rarely it is
misinterpreted as  saying that the thermal bath experienced by accelerated 
observers in the vacuum state of a quantum field would be  indistinguishable from a thermal state of the same field 
at  temperature $T = T_{\rm U}$ according to  inertial observers  --- which is a  false statement. This overly stringent
view has compelled many to challenge
or restrict the validity of the Unruh effect based on non-local observables
which behave differently in these two situations. Simple examples of such observables
are two-point correlation functions\cite{AS}. Also, the fact that uniformly accelerated observers,
static with respect to each other, can have different proper accelerations $a$ (depending on their separation), 
makes  the Unruh temperature $T_{\rm U}$ spatially inhomogeneous across the uniformly accelerated frame. This peculiar behavior
has been held against the interpretation of $T_{\rm U}$ 
as a physical temperature when spatially-extended  systems are considered\cite{BV}.
Although having no
analogue for an inertial (equilibrium) thermal 
state, this inhomogeneity of $T_U$  is mandatory in the accelerated frame. This is a direct, well-known consequence
of relativistic redshift effects\cite{Tolman}, as we shall discuss later.

Our main goal here is to settle this debate by
providing convincing evidence that the strict thermal nature of the Minkowski vacuum in the uniformly accelerated frame
is physically meaningful also for spatially-extended systems --- which sense non-local observables
and inhomogeneous  $T_{\rm U}$ ---, 
upholding $T_{\rm U}$ as a legitimate temperature. In order to achieve this, we analyze  a
system composed of two uniformly accelerated spins, separated by an arbitrary finite distance $d$ (fixed in their accelerated rest frame),
 directly coupled to each  other --- which confers unity to the system ---, and locally coupled to  quantum  fields in the vacuum state.
We focus attention on the  reduced density matrix of the accelerated-spins' system and show that it evolves to an 
equilibrium state which, according to an arbitrary observer with proper acceleration $a$ static with the spins, 
is exactly the one which would be expected if the system were in contact with a thermal bath with 
temperature $T=T_{\rm U}$; in other words, we show that the spin system  thermalizes at a nonzero, well-defined temperature 
due to the vacuum fluctuations it experiences in its accelerated rest frame. This corroborates 
the view that although one might construct observables which distinguish the Unruh thermal bath
from an ordinary (i.e., inertial) one at the same temperature --- and, in fact, we show that the thermalization time scales can distinguish
these two situations, which, we stress, is  not in conflict with the Unruh effect ---,  
the former does act as a legitimate thermal reservoir also for extended systems.
    
\section{Results}

\subsection{The setup.}

Unless stated otherwise, we adopt
natural units, in which $\hbar = c = k_{\rm B} = 1$.
Let us consider two  spin-$1/2$ point particles, A and B, whose spins $\hat{\bf{s}}_{\rm A} $ and $\hat{\bf{s}}_{\rm B}$ are directly coupled 
to each other  via, say, the (free) Hamiltonian $\hat H_0 = -J \hat{s}_{\rm A}^{\rm z} \hat{s}^{\rm z}_{\rm B}$, $J \neq 0$, which has two-fold degenerate eigenvalues 
$\pm J/4$.  Since the spins are taken to be spatially separated, this is a simple, yet legitimate model of an extended system. 
Now, let us couple (locally and weakly) the spins to a quantum  field, such that neither an eventual 
constant of motion would prevent the spin system from thermalizing, nor the dynamics could
be mapped onto a two free-particle problem. As a matter of fact, the simplest spin-field interaction which would lead to some interesting evolution 
is given by linearly coupling one of the other spin components, say $\hat s_{\rm A(B)}^{\rm x}$, to
a massless, scalar quantum field $\hat \Phi$. However, this would lead to a conservation law
for the observable $\hat s_{\rm A}^{\rm x} \hat s_{\rm B}^{\rm x}$, which, in turn, would split the state space of the spin system as if it were two noninteracting (non-localized) spins.
In order to avoid such a symmetry, we shall couple the other spins' component, $\hat s_{\rm A(B)}^{\rm y}$, to another massless, scalar field, so that the (time-dependent) 
Hamiltonian is given by:
\begin{eqnarray}
\hat{H} (\tau)= -J \hat{s}_{\rm A}^{\rm z} \hat{s}^{\rm z}_{\rm B} +q\sum_{j\in\{{\rm x},{\rm y}\}}\left[\frac{\hat{\Phi}^j_{\rm A}(\tau) \hat{s}^j_{\rm A}}{u_{\rm A}^0(\tau)}+\frac{\hat{\Phi}^j_{\rm B}(\tau) \hat{s}^j_{\rm B}}{u_{\rm B}^0(\tau)}\right],
\label{H}
\end{eqnarray}
where $q \in {\mathbb R}$ is a dimensionless (scalar) coupling constant, 
$\hat{\Phi}^j_{\rm A(B)}(\tau) := \hat{\Phi}^j(\tau,\bf{x}_{\rm A(B)}(\tau))$ are 
massless, scalar quantum fields --- which are independent for
different $j\in \{{\rm x},{\rm y}\}$ --- evaluated at the spins' location $\bf{x}_{\rm A(B)}(\tau)$, and
$u^0_{\rm A(B)}:= d\tau/d\tau_{\rm A(B)}$ is the time component of the four-velocity of spin A(B), with $\tau_{\rm A(B)}$ being its proper time.
Note that there is no 
need to include the free Hamiltonian of the fields $\hat \Phi^j$ in Eq.~(\ref{H})  since we already take into account
the explicit time-dependence of $\hat \Phi^j$ enforced by the free-field Klein-Gordon equation, $\square \hat \Phi^j = 0$.
The coordinate system $\{(\tau,\bf{x})\}$ would be arbitrary at this point.
However, since we are going to consider that the system evolves according to the von Neumann equation (Eq.~(\ref{drhodt}) below), 
it is necessary that $\tau$ represents the parameter of a time-translation symmetry of the spacetime --- $\partial /\partial \tau$ is a time-like Killing field --- and that 
$\hat H$ given in Eq.~(\ref{H}) is the  Hamiltonian of the system in the (stationary) reference frame associated with this symmetry. 
For instance, if the spins  were  static in an inertial frame, then  $\tau$ could be conveniently set to be the usual inertial time --- for which $u_{\rm A(B)}^0 = 1$. For
accelerated spins which are static in a uniformly accelerated frame, as we are interested here,  $\tau$ 
can be interpreted as the proper time of a fiducial uniformly accelerated observer with respect to (w.r.t.)~whom the spins are static.
The presence of $u_{\rm A(B)}^0$ in the interaction terms accounts for the fact that  $\hat H$ evolves the system in the time parameter 
$\tau$, while
$q \,\hat \Phi_{\rm A(B)} {\hat s}^j_{\rm A(B)}$, being a local interaction, is related to the evolution in the time parameter $\tau_{\rm A(B)}$.
In particular, care must be taken when interpreting
the meaning of the parameter $J$: it is twice the energy gap $\Delta E$ of the spin system  as measured by the fiducial observer.
According to an observer at spin-A(B) location, this gap is inevitably corrected by the ``redshift'' factor $u_{\rm A(B)}^0$: $\Delta E_{\rm A(B)} = u_{\rm A(B)}^0 \Delta E$.
Without any loss of generality, 
the fiducial observer can be placed at, say, spin-A's position, so that $\tau = \tau_{\rm A}$ 
--- which leads to
$u_{\rm A}^0 = 1$. As we shall see later,
the value of $u_{\rm B}^0$ depends on whether the acceleration of the system is perpendicular or parallel to  its spatial 
extension.
(The fact that $\Delta E_{\rm A}/\Delta E_{\rm B} = u_{\rm A}^0/u_{\rm B}^0$ --- which is independent of $J$ and of the choice of fiducial observer ---
plays an essencial role in interpreting our results in the parallel case,  in which our extended system feels an
inhomogeneous Unruh temperature.)
The  model defined 
by Eq.~(\ref{H}) can be considered as an extension of the well-known spin-boson model, which is taken as a paradigm for the study of the dissipative dynamics of two-level systems\cite{leggett}.

Let
$\hat{\rho}$ be the positive semidefinite, Hermitian, trace-class operator (with trace 1)  describing the state of the whole universe (spins $+$ fields). Its evolution is governed by 
\begin{eqnarray}
{\rm i}\partial_\tau \hat{\rho} = [\hat{H}(\tau),\hat{\rho}],
\label{drhodt}
\end{eqnarray}
whose solution can be written as
\begin{eqnarray}
\hat{\rho}(\tau) = \hat{U}(\tau,\tau_0) \hat{\rho}_0 [\hat{U}(\tau,\tau_0)]^{-1},
\label{rhoevol}
\end{eqnarray}
with $\hat{\rho}_0 := \hat{\rho}(\tau_0)$ and $\hat{U}(\tau,\tau_0)$ satisfying
\begin{eqnarray}
{\rm i}\partial_\tau\hat{U}(\tau,\tau_0)= \hat{H}(\tau) \hat{U}(\tau,\tau_0), \;\;\;\hat{U}(\tau_0,\tau_0)=\hat 1. 
\label{U}
\end{eqnarray}
We are only interested in the reduced density matrix of the spin system, obtained after tracing out the fields' degrees of freedom (system's reduced matrix):
\begin{eqnarray}
\hat{\rho_{\rm s}}(\tau) := {\rm tr}_\Phi \left[\hat{\rho}(\tau) \right].
\label{rhos}
\end{eqnarray}

Motived by the results obtained from the spin-boson model\cite{leggett}, we shall treat the coupling with the quantum fields as a (time-dependent) 
perturbation, $ \hat V(\tau) :=q\sum_{j\in\{{\rm x},{\rm y}\}}  [\hat{\Phi}^j_{\rm A}(\tau) \hat{s}^j_{\rm A}/u_{\rm A}^0(\tau)+\hat{\Phi}^j_{\rm B}(\tau) \hat{s}^j_{\rm B}/u_{\rm B}^0(\tau)]$, 
on the free Hamiltonian $\hat{H}_0 = -J \hat{s}_{\rm A}^{\rm z} \hat{s}^{\rm z}_{\rm B}$. Indeed, under this regime, namely, the weak coupling regime, the spin-boson model provides means for observing spin thermalization process, with predicted decoherence/relaxation time scales matching those observed in physical systems satisfying the conditions imposed.  
For that, we write $\hat U(\tau,\tau_0) = {\rm e}^{-{\rm i}\hat H_0 (\tau-\tau_0)} \hat U_{\rm I}(\tau,\tau_0)$, where $\hat U_{\rm I}$ represents the time evolution operator in the interaction picture, satisfying
\begin{eqnarray}
{\rm i}\partial_\tau\hat{U}_{\rm I}(\tau,\tau_0)= \hat{H}_{\rm I}(\tau) \hat{U}_{\rm I}(\tau,\tau_0),\;\;\;\hat{U}_{\rm I}(\tau_0,\tau_0)= \hat 1,
\label{dUIdt}
\end{eqnarray}
with
\begin{eqnarray}
\hat{H}_{\rm I}(\tau) :={\rm e}^{{\rm i}\hat H_0 \Delta \tau} \hat V(\tau) {\rm e}^{-{\rm i}\hat H_0 \Delta \tau}
=q \sum_{j\in\{{\rm x},{\rm y}\}}\left[\frac{\hat{\Phi}^j_{\rm A}(\tau) \hat{s}^j_{\rm A}(\tau)}{u_{\rm A}^0(\tau)}+\frac{\hat{\Phi}^j_{\rm B}(\tau) \hat{s}^j_{\rm B}(\tau)}{u_{\rm B}^0(\tau)} \right],
\label{HI}
\end{eqnarray}
where  $\Delta \tau := \tau -\tau_0$ and
\begin{eqnarray}
 \hat s^j_{M}(\tau)
=\hat s^j_{M} \cos\left(\frac{J \Delta\tau}{2}\right)+2{\rm i}\left[ \hat s^j_{M},  \hat s^{\rm z}_{M}\right]\hat s^{\rm z}_{\bar {M}} \sin\left(\frac{J \Delta \tau}{2}\right),
\label{Had}
\end{eqnarray}
with $M\in \{{\rm A},{\rm B}\}$ and $\bar {\rm A} := {\rm B}$, $\bar {\rm B} := {\rm A}$.

Solving Eq.~(\ref{dUIdt}) iteratively (as a Dyson series),
working consistently up to second order in $q$, and  restricting attention to the case where the initial state is simply separable, $\hat\rho_0 = \hat\rho_{\rm s0}\otimes \hat\rho_{\Phi 0}$, where
$\hat\rho_{\rm s0}$ and   $\hat\rho_{\Phi 0}$ describe the initial state of the spin system and of the fields, respectively, we obtain, from Eq.~(\ref{rhos}): 
\begin{eqnarray}
\hat \rho_{\rm s}(\tau) &=& {\rm e}^{-{\rm i}\hat H_0 \Delta \tau}\Big\{\hat \rho_{\rm s0}
-\frac{q^2}{2}\sum_{
\tiny{{M},{N}\in\{{\rm A},{\rm B}\}}
}
\int_{\tau_0}^\tau \frac{d\tau'}{u_{M}^0(\tau')} \int_{\tau_0}^\tau \frac{d\tau''}{u_{N}^0(\tau '')} 
\sum_{{ j}\in\{{\rm x},{\rm y}\}}
{\rm i} G^j_{\rm F}(x_{M}',x_{N}'')
\nonumber \\
& &\hskip 1.5cm
{\rm T}\left\{\left[\hat s_{M}^j(\tau'),
\hat s_{N}^j(\tau'')\hat \rho_{\rm s0} \right]\right\}
+{\rm H.c.}\Big\}\,{\rm e}^{{\rm i}\hat H_0 \Delta \tau}
+ {\cal O} (q^3),
 \label{rhos3}
\end{eqnarray}
where H.c.\ stands for the Hermitian conjugate of the term which precedes it
and ${\rm i}G^j_{\rm F}(x',x'') := {\rm tr}_\Phi\left\{\hat \rho_{\Phi 0}{\rm T}\left[\hat \Phi^j (x') \hat \Phi^j(x'')\right]\right\}$ are the 
time-ordered Feynman correlators  in
state $\hat \rho_{\Phi 0}$.  (The 
usual time-ordering operator ${\rm T}$ appearing explicitly in the second line of Eq.~(\ref{rhos3}) must be applied before
 the commutator is expanded.) Since we are interested only in the effects of quantum fluctuations of $\hat \Phi^j$ on the spin system, we 
have already assumed 
$\langle\Phi^j(x)\rangle := {\rm tr}_\Phi\left\{\hat \rho_{\Phi0} \hat\Phi^j(x)\right\} = 0$, which, together with the independence of $\hat \Phi^j$ for different $j$, implies
${\rm tr}_\Phi\left\{\hat \rho_{\Phi 0}\hat \Phi^{\rm x} (x') \hat \Phi^{\rm y}(x'')\right\}=0$. Also, we restrict attention to the case where
 ${\rm i}G^{\rm x}_{\rm F}(x',x'') \equiv {\rm i}G^{\rm y}_{\rm F}(x',x'')=:{\rm i}G_{\rm F}(x',x'')$, which applies to the situation  in which  we are most interested.

\subsection{Static spins' arrangements in static field states.}

Restricting attention to static spins' arrangements $\bf{x}_{\rm A}$, $\bf{x}_{\rm B}$  and  static field states $\hat \rho_{\Phi 0}$ 
(w.r.t.~the time parameter $\tau$), 
it follows that $G_{\rm F}(x_{M}',x_{N}'')$ can depend on $\tau'$ and $\tau''$ only through the combination $\xi := \tau'-\tau''$, $G_{\rm F}(x_{M}',x_{N}'')=:G_{{MN}}(\xi)$ --- in addition to
$u_{M}^0(\tau) \equiv u_{M}^0$ being constant.
This suggests that it may be more convenient, in the 
second-order term of Eq.~(\ref{rhos3}), to perform a change of integration variables to
$\eta:=( \tau'+\tau'')/2$ and $\xi$. 
 Notice that, by construction,  $G_{{MN}}(\xi)=G_{NM}(-\xi)$, which, in particular, implies that
$G_{\rm AA}(\xi)$ and $G_{\rm BB}(\xi)$ are even distributions w.r.t.~$\xi$.  But staticity 
also implies that $G_{\rm AB}(\xi)$ and $G_{\rm BA}(\xi)$ are even distributions w.r.t.\ $\xi$; hence,
$G_{\rm AB}(\xi)\equiv G_{\rm BA}(\xi)$.
Using Eq.~(\ref{Had}) into Eq.~(\ref{rhos3}), the integral in $\eta$ can be 
 explicitly evaluated, leading to:
\begin{eqnarray}
{\rm e}^{{\rm i}\hat H_0 \Delta \tau}\hat \rho_{\rm s}(\tau){\rm e}^{-{\rm i}\hat H_0 \Delta \tau}&=& \hat \rho_{\rm s0} 
-\frac{q^2}{2}
\sum_{
\tiny{{M},{N}\in\{{\rm A},{\rm B}\}}}
\int_{-\Delta \tau}^{\Delta \tau} d\xi\,  \frac{{\rm i} G_{MN}(\xi)}{u_{M}^0 u_{N}^0} 
\left\{
\left(\Delta\tau -|\xi |\right)\left[\hat C^{(+)}_{(MN)}\cos\left(\frac{J\xi}{2}\right)
\right.
\right. \nonumber \\
& &
\left.  +\hat D^{(-)}_{(MN)}\sin\left(\frac{J|\xi |}{2}\right)\right]
 +
\frac{2}{J}\sin\left(\frac{J(\Delta\tau -|\xi |)}{2}\right)\left[\hat C^{(-)}_{(MN)}\cos\left(\frac{J\Delta\tau}{2}\right)
\right.
\nonumber \\
& &\left. \left.
+\hat D^{(+)}_{(MN)}\sin\left(\frac{J\Delta\tau}{2}\right)\right]
\right\}
+{\rm H.c.}
 + {\cal O} (q^3),\;\;
 \label{rhogenfinal}
\end{eqnarray}
where 
\begin{eqnarray}
\hat C^{(\pm)}_{MN} := \sum_{\tiny{{j}\in \{{\rm x},{\rm y}\}}}\left\{\frac{1}{2}\left[\hat s^j_{M},\hat s^j_{N} \hat \rho_{s0}\right]\pm 2\left[\hat s^{\bar j}_M \hat s^{\rm z}_{\bar M},\hat s^{\bar j}_N \hat s^{\rm z}_{\bar N} \hat \rho_{\rm s0}\right] \right\},
\label{Cpm}\\
\hat D^{(\pm)}_{MN} := \sum_{\tiny{j\in \{{\rm x},{\rm y}\}}}\epsilon_{j \bar j}\left\{\left[\hat s^{\bar j}_M \hat s^{\rm z}_{\bar M},\hat s^{j}_N\hat \rho_{\rm s0}\right]\pm \left[\hat s^{j}_M ,\hat s^{\bar j}_N \hat s^{\rm z}_{\bar N} \hat \rho_{\rm s0}\right] \right\},
\label{Dpm}
\end{eqnarray}
with $\bar {\rm x} := {\rm y}$, $\bar {\rm y} := {\rm x}$, $\epsilon_{\rm xy}=-\epsilon_{\rm yx}= 1$, and indices $M,N$ inside
parentheses in Eq.~(\ref{rhogenfinal}) 
denote symmetrization: $X_{(MN)}:= (X_{MN}+X_{NM})/2$. 

As it stands, Eq.~(\ref{rhogenfinal}), being a truncated perturbative expansion, 
is not appropriate to investigate long-term features of the spin system,
as relaxation to an eventual equilibrium state when $\Delta \tau \to \infty$. In this limit,
the second-order term in $q$ is, in general, unbounded and, therefore, cannot be consistently considered as providing
a ``small'' deviation from the free evolution. We can, nonetheless, try to break long-term evolution 
into a sequence 
of $N$~($\gg 1$) time lapses $\Delta \tau$ such that in each time lapse, for sufficiently small coupling~$q$, 
the spins' evolution is
well described by Eq.~(\ref{rhogenfinal}). 
This strategy is trivially valid for closed systems. Here, however, since
tracing out the fields' degrees of freedom at the end of  each  time step does not necessarily lead to the same
result as taking the trace only after $N$ steps,  this procedure is not guaranteed, in general, to lead to the correct evolution of the reduced density matrix.
Notwithstanding this, 
in the Methods' subsection ``Validity of the Markovian regime,'' we show that there is a finite time-lapse scale $\Delta \tau$ ($ \gg J^{-1},\|{\bf x}_{\rm A}-{\bf x}_{\rm B}\|$)
for which this strategy holds true --- resembling a Markovian regime ---, leading to the long-term evolution
\begin{eqnarray}
\hat\rho_{\rm s}(\tau_N) 
= {\rm e}^{-{\rm i} \hat H_0 \tau_N }\left({\rm e}^{-q^2{\rm R}_0 \tau_N}\hat\rho_{\rm s0}\right){\rm e}^{{\rm i} \hat H_0 \tau_N},
\label{evolfinal}
\end{eqnarray}
where $\tau_N = N\Delta\tau$ 
and ${\rm R}_0:{\cal T}\left({\cal H}_{\rm s}\right)\to {\cal T}\left({\cal H}_{\rm s}\right)$  --- 
an operator acting on the space of trace-class operators describing the spin system --- is
determined by the equation [recall Eqs.~({\ref{Cpm}) and (\ref{Dpm})]
\begin{eqnarray}
{\rm R}_0 \left(\hat\rho_{{\rm s}}\right)= \frac{1}{2}
\sum_{
{M,N\in\{{\rm A},{\rm B}\}}}
 \left.\left\{ {\rm i} \pi\widetilde{G}_{MN}(J/2)\, \hat C^{(+)}_{(MN)} -{\cal P}_{J/2}\left[{\rm i}\widetilde{G}_{MN}\right]\hat D^{(-)}_{(MN)} 
\right\}\right|_{\hat\rho_{{\rm s}0}\mapsto \hat\rho_{{\rm s}}}
+{\rm H.c.},
 \label{R0}
\end{eqnarray}
with
\begin{eqnarray}
\widetilde{G}_{MN}(\omega) := \frac{1}{2\pi}\int_{-\infty}^\infty d\xi \,\frac{{G}_{MN}(\xi)}{u_M^0 u_N^0} \,{\rm e}^{{\rm i}\omega \xi}
\label{Gftilde}
\end{eqnarray}
and
\begin{eqnarray}
{\cal P}_a[f] := \lim_{\epsilon \to 0}\int_{{\mathbb R} \backslash [-\epsilon,\epsilon]}\frac{f(x+a)}{x}\,dx.
\label{PV}
\end{eqnarray}

In the Methods' subsection ``Decay modes and related decay rates of the spin system,'' we present  all
eigenvalues $\lambda_k$ and (right) eigenvectors (or ``eigenmatrices'') $\hat \rho_k$ of ${\rm R}_0$ (with $k=1,\dots,16$) in terms of ${\rm i}\widetilde{G}_{MN}(J/2)$, 
${\cal P}_{J/2}\left[{\rm i}\widetilde{G}_{MN}\right]$, and the elements of the Bell basis $\{|\Psi_{\rm AB}^{(\pm)}\rangle,|\Phi_{\rm AB}^{(\pm)}\rangle\}$ defined in Eqs.~(\ref{Bellbasis1})
and (\ref{Bellbasis2})  --- see Eqs.~(\ref{alphas}-\ref{rho16}); note that each individual mode $\hat{\rho}_k$ does not necessarily have to represent a physical state.
This encodes complete information about the evolution of the 
spin system.
It is a simple task to verify
that these eigenmatrices $\hat\rho_k$ are also eigenmatrices of the free
evolution: ${\rm E}_0(\hat\rho_k) := {\rm e}^{-{\rm i}\hat H_0 \Delta \tau}\hat \rho_k {\rm e}^{{\rm i} \hat H_0 \Delta \tau} = {\rm e}^{-{\rm i} E_k \Delta\tau} \hat \rho_k$, where,
depending on $k$,
$E_k = 0,\pm J/2$. Therefore, noticing that $\lambda_1=E_1 = 0$ and provided 
${\rm Re}(\lambda_{k\neq 1}) > 0$ --- as will be verified later in our cases of interest ---, we finally obtain the evolution of the spin system in the Markovian regime:
\begin{eqnarray}
\hat\rho_{\rm s}(\tau_N) =\sum_{k=1}^{16} c_k {\rm e}^{-(q^2\lambda_k+{\rm i}E_k) \tau_N}\hat\rho_{k} 
\toover[2]{N\to\infty}
\hat\rho_1 =: \hat \rho_{\rm eq},
\label{evolfinalasymp}
\end{eqnarray}
where the coefficients $c_k$ are uniquely determined by the initial condition $\sum_{k=1}^{16} c_k\hat\rho_k = \hat\rho_{\rm s0}$ --- in particular, $c_1 = {\rm tr}_{\rm s}\left\{\hat\rho_{\rm s0}\right\} = 1$.

\subsection{Uniformly accelerated spins in the vacuum.}
\label{sec:uas}

Finally, in this subsection we apply the  expressions obtained above to our case of interest: uniformly accelerated spins in the vacuum.
The vacuum state  $|0\rangle$ of a free, massless scalar field is characterized as being the (unique) Poincar\'e-invariant state of the theory.
The vacuum  expectation value of the field vanishes,  $\langle 0|\hat\Phi(x)|0\rangle = 0$, whereas
its two-point (Wightman) function is given by 
\begin{eqnarray}
W(x,x') := \langle 0 |\hat \Phi(x) \hat \Phi(x')|0\rangle =
\frac{1}{4 \pi^2\sigma_\epsilon(x,x')},
\label{W}
\end{eqnarray} 
where 
$\sigma_\epsilon(x,x')$ is the ($\epsilon$-regularized) square of the geodesic ``distance''
between events $x$ and $x'$ --- which is obtained from the square of the geodesic distance, $\sigma (x,x')$, by
introducing an infinitesimal  negative imaginary part ($-{\rm i}\epsilon$) into the time coordinate of the first event $x$.
As expected from  its definition, notice that $W(x,x')$ is a bi-scalar. Therefore, its value is insensitive to the choice of coordinate 
system we use to represent
the events $x$ and $x'$.
In terms of inertial Cartesian coordinates $\{(t,X,Y,Z)\}$, $\sigma(x,x') = -(t-t')^2+(X-X')^2 +(Y-Y')^2+(Z-Z')^2$, whereas in terms of 
coordinates $\{(\tau,X,Y,\zeta)\}$ well adapted to
a uniformly accelerated frame --- 
defined through
 $t = (\zeta + a^{-1}) \sinh(a\tau)$, $Z = (\zeta + a^{-1}) \cosh(a\tau)$, with $\zeta > - a^{-1}$, $\tau \in {\mathbb R}$ ---, we have
\begin{eqnarray}
\sigma(x,x') = -\frac{4}{a^2} (a\zeta+1)(a\zeta'+1) \left[\sinh\left(\frac{a(\tau-\tau')}{2}\right)\right]^2
+ (X-X')^2+(Y-Y')^2 +(\zeta - \zeta')^2,
\label{sigma2}
\end{eqnarray}
where $a>0$ is a constant.
The interpretation of  $\tau$ and $\zeta$ follows from the form of the Minkowski line element in these coordinates,
\begin{eqnarray}
ds^2&=&
-(1+a\zeta)^2 d\tau^2+dX^2+dY^2+d\zeta^2:
\label{Rle}
\end{eqnarray}
the coordinate $\tau$ is the proper time of (fiducial) observers static at $\zeta = 0$ ---
whose constant proper acceleration is given by the parameter  $a$ ---, whereas 
the coordinate $\zeta$ measures spatial distances along the acceleration direction, according to observers static in this coordinate system.
Notice [for the sake of the discussion below Eq.~(\ref{H})] that $\tau$ does represent a time-translation symmetry of the spacetime. This can be inferred from the line element
given in Eq.~(\ref{Rle}), since the coefficients of the differentials (i.e., the  metric components) are independent of $\tau$ --- $\partial/\partial \tau$ is
a time-like Killing field known as the  boost Killing field.

For the sake of completeness, let us recall some basic facts about accelerated frames which are relevant for our purposes. Fig.~\ref{fig:rw} presents a diagram 
depicting the Minkowski spacetime region  $Z>|t|$ (called  Rindler wedge) which is covered by the accelerated Cartesian coordinates $\{(\tau,X,Y,\zeta)\}$
defined above  Eq.~(\ref{sigma2}) (suppressing
the $X$ and $Y$ directions). In such a diagram, worldlines of inertial observers would be represented by straight lines making an angle smaller than $45^{\rm o}$ with the vertical
$t$ axis, while light rays are represented by $45^{\rm o}$ lines. Constant-$\zeta$ (hyperbolic) 
curves represent worldlines of a family of uniformly-accelerated observers ``at rest'' w.r.t.\ each other --- which constitutes a uniformly-accelerated frame. 
All these hyperbolas (for given $X$ and $Y$) asymptote the same light rays (on ${\cal H}^+$ and ${\cal H}^-$) 
which intersect at $S$  (a flat $2$-surface
when  $X$ and $Y$ are restored). This shows that, contrary to inertial frames, no single  uniformly-accelerated frame can cover the whole (Minkowski)  spacetime. Moreover, to any 
uniformly-accelerated observer,
there is a causally inaccessible spacetime region, with ${\cal H}^+$ playing the role of an  event horizon.  This particular feature is essencial
in understanding how a pure quantum state (the Minkowski vacuum) is perceived  as a  mixed (thermal) state by uniformly-accelerated observers. In terminology introduced in Ref.\cite{Espagnat}, the Unruh thermal bath
is an example of an  improper  mixed state, arising from tracing out some degrees of freedom 
of a pure state --- the ones inaccessible to the accelerated observers. In contrast, inertial thermal states are
truly statistical  mixtures of pure states (energy-momentum eigenstates according to inertial observers), 
representing the whole system ---
which qualifies them as  proper
mixed states. Although this  kind of distinction may be relevant  on a conceptual level
--- see, e.g., Ref.\cite{MOPS} for a discussion in the context of information loss
in black hole evaporation ---,
it plays no role for our purposes since observables restricted to the Rindler wedge cannot 
uncover the improper nature of the Unruh thermal bath.

\begin{figure}[t]
\includegraphics[scale=0.90]{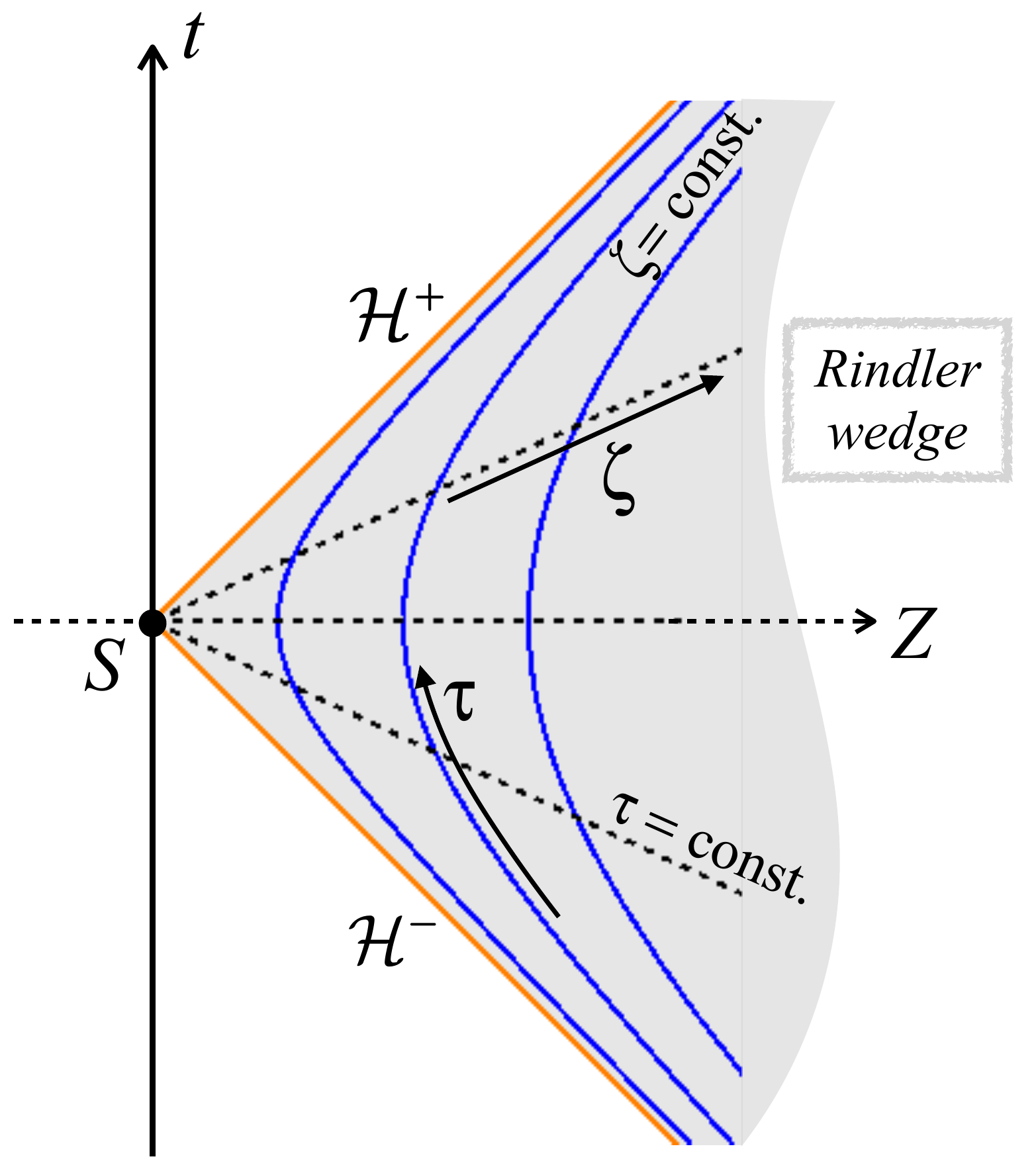}
\caption{Uniformly accelerated frame and the Rindler wedge. Depiction of the Minkowski spacetime region $Z > |t|$ covered by the coordinates $\tau$ and $\zeta$ (suppressing coordinates $X$ and $Y$). 
This is commonly referred to as the  Rindler wedge. Solid (blue)  curves represent the ($\zeta = $~constant) worldlines of uniformly-accelerated observers 
``at rest'' w.r.t.\ each other, while
dashed (black) lines represent $\tau =$~constant hypersurfaces --- which encodes simultaneity according to
these observers. All constant-$\zeta$ curves asymptote ${\cal H}^+$ and ${\cal H}^-$, reflecting  that 
this uniformly-accelerated frame does not cover the whole spacetime. Moreover, there is a spacetime region which is
causally
inaccessible to observers in this frame, for whom ${\cal H}^+$ represents a (future) event horizon.}
\label{fig:rw}
\end{figure}

The fact that constant-$\zeta$ curves have common asymptotes --- instead of simply  being translated hyperbolas in the $Z$ direction --- also shows that uniformly-accelerated observers static at
different values of  $\zeta $ have different proper accelerations. In fact, it can be shown that $\zeta =$~constant worldlines have proper acceleration given 
by ${a}_\zeta = a/(1+a\zeta)$ --- and it follows directly from Eq.~(\ref{Rle}) that observers following these worldlines have proper time given by ${\tau}_\zeta = (1+ a\zeta)\tau$.
Thus, according to the Unruh effect,
each such observer ``feels'' a different Unruh temperature 
$T_{\rm U} = {a}_\zeta/(2\pi)$. More concretely, a uniformly-accelerated extended system (with constant  proper spatial dimensions)
would ``feel'' an inhomogeneous Unruh temperature along the direction of its acceleration. This inhomogeneity --- which our simple model for an extended system can probe ---
 is at the heart of the arguments against $T_{\rm U}$ being a
legitimate 
physical temperature\cite{BV}. Note, however, that $T_{\rm U} \sqrt{-g_{00}} =$~constant --- $g_{00}$ being the time-time metric component read from Eq.~(\ref{Rle}) ---,
evidencing the role played by the redshift effect in the spatial variation of $T_{\rm U}$. This constraint, 
$T \sqrt{-g_{00}} =$~constant, is known as Tolman's relation\cite{Tolman} and is valid for thermal states in arbitrary (flat and curved) stationary spacetimes --- for
it
follows simply from the condition that the net heat flow in an equilibrium state must
vanish everywhere.

It is worth pointing out that all this discussion about uniformly-accelerated frames, event horizons, 
and the Unruh effect is only relevant for
 interpreting our final results, not for carrying out any of the calculations. As far as the calculations are concerned,
all we need is the vacuum two-point function $W(x,x')$, given by Eq.~(\ref{W}), and information about the spins' worldlines. 
The use of $\tau$ and $\zeta$ instead of 
$t$ and $Z$ to express the bi-scalar $\sigma(x,x')$ --- in Eq.~(\ref{sigma2}) ---  can be seen as a mere mathematical convenience, 
since uniformly-accelerated worldlines separated by a constant proper distance take the simple form 
$\zeta =$~constant.
In addition, nowhere in the calculations we express the (pure) Minkowski vacuum 
state as a  thermal bath of (Rindler) particles according 
to uniformly accelerated observers --- i.e., we do not  assume the Unruh effect to be true. From our perspective, we are inertial theorists using standard quantum theory to
predict the behavior of an extended ``thermometer'' accelerating in the vacuum.

\subsection{Spins with equal proper accelerations.}

Let us first consider the case where the spins are uniformly accelerated  perpendicularly to their spatial separation, with the same proper acceleration $a$. This can be described,  in the  coordinates 
$\{(\tau, X,Y,\zeta)\}$, by spin trajectories given by
$X_{\rm A}(\tau) \equiv Y_{\rm A}(\tau)\equiv \zeta_{\rm A}(\tau)\equiv Y_{\rm B}(\tau)  \equiv \zeta_{\rm B}(\tau) \equiv0$, $X_{\rm B} (\tau)\equiv d$,
where $d$ is the spatial separation between the spins (in their rest frame) and, conveniently,  $\tau \equiv \tau_{\rm A} \equiv \tau_{\rm B}$ (i.e., $u_{\rm A}^0 = u_{\rm B}^0 = 1$).
This, combined with the Wightman function given in Eq.~(\ref{W}), is all we need to calculate the eigenvalues
$\lambda_k$ and eigenmatrices $\hat\rho_k$ appearing in  the long-term 
evolution of the spin system [Eq.~(\ref{evolfinalasymp})], as presented in more detail in the
Methods' subsection ``Transformed Feynman correlators and their Principal Values.''

The nonzero eigenvalues $\lambda_k$ ($k=2,\dots,16$) are related, through
Eq.~(\ref{evolfinalasymp}), to the decoherence/relaxation rates of the spin system. In Fig.~\ref{ratesd}, we plot all  these rates (normalized by $q^2$) as functions of the proper acceleration $a$,
for different separations $d$.
\begin{figure*}
 \includegraphics[width=\textwidth,height=11cm]{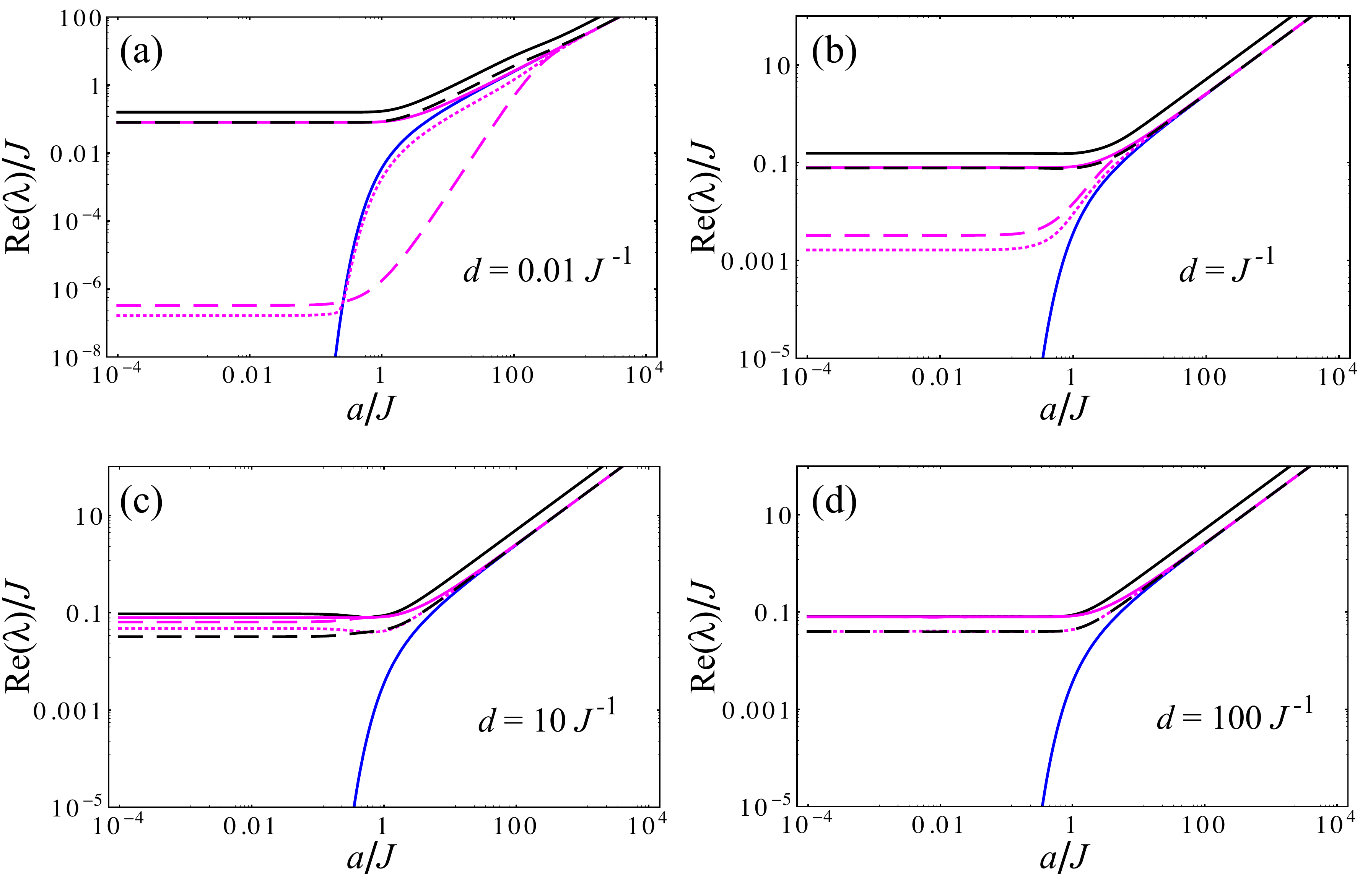}
\caption{Acceleration dependence of relaxation/decoherence rates of the spin system with equal proper accelerations.
We plot the decay rates (normalized by $q^2$),  ${\rm Re}(\lambda_k)$ --- $k = 2$ (solid, black line), $ 3$ (dashed, magenta line),
$4,5,6$ (solid, blue line),
$7,8,9,10$ (dotted, magenta line), $11,12,13,14$ (dashed, black line), $15,16$ 
(solid, magenta line)
---, of the decaying modes of the spin system,
as functions of the spins' acceleration $a$, for different separations $d$. 
 Unless $d\ll J^{-1}$ --- in which case mode
$\hat\rho_3$ dominates the late-time dynamics for $J \lesssim a \lesssim d^{-1}$ ---, modes $\hat\rho_4,\hat\rho_5,\hat\rho_6$  (all
belonging to the lowest-energy subspace)
dictate
how the system approaches equilibrium with a relaxation/decoherence rate 
given by $q^2 J/\left[8\pi \left({\rm e}^{\pi J/a}-1\right)\right]$.}
\label{ratesd}
\end{figure*}
In case $d \gtrsim J^{-1}$, there are basically two regimes of acceleration: $a\ll J$ --- in which modes 
$\hat\rho_4,\hat\rho_5,\hat\rho_6$  (all
belonging to the lowest-energy subspace)
dictate
how the system approaches equilibrium with a relaxation/decoherence rate 
given by $q^2 J{\rm e}^{-\pi J/a}/(8\pi)$ --- and $a\gg J$ --- in which all relaxation/decoherence rates
degenerate in just two values: $q^2 a/(8\pi^2)$ and twice this value. Such a result, namely, a relaxation rate that scales as a power law of the temperature --- recall that $T_{\rm U} = a/(2\pi)$ --- 
is known to be a signature of certain inertial baths in the limit of high temperatures. Indeed, for the class of baths known as Ohmic environments, it is precisely established for the spin-boson model that the decay rate will have a power law dependence, which is linear in a second order system-bath coupling perturbation theory\cite{leggett}.
In case $d \ll J^{-1}$, there appears a third, moderate regime of acceleration ($J\lesssim a \lesssim d^{-1}$)
in which mode $\hat\rho_3$ dictates how equilibrium is approached, with a relaxation/decoherence rate given approximately by
$q^2 J d^2 a^2/(12 \pi)$.
\begin{figure*}
 \includegraphics[width=\textwidth,height=11cm]{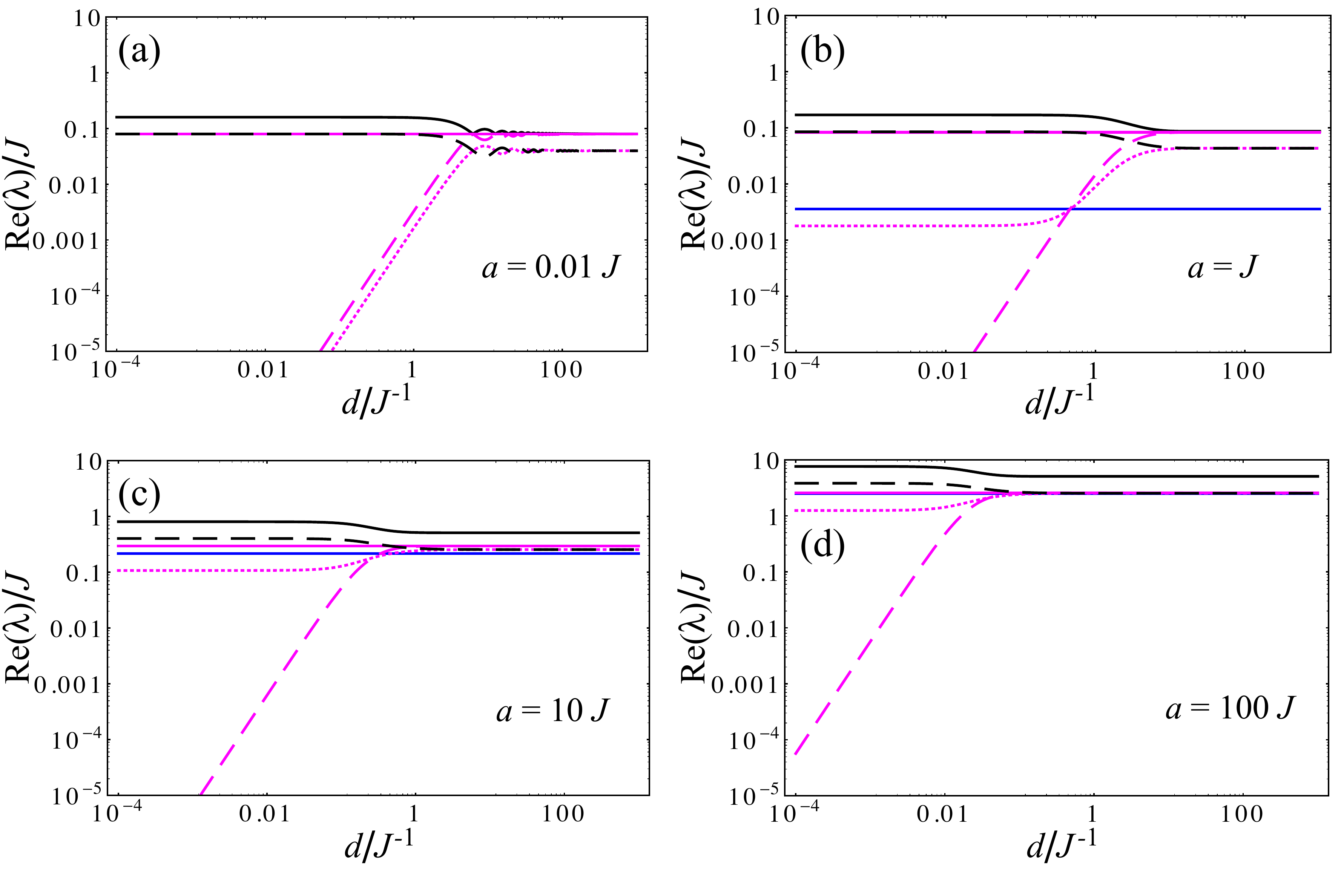}
\caption{Spin-separation dependence 
of relaxation/decoherence rates of the spin system with equal proper accelerations.
We plot the decay rates (normalized by $q^2$),  ${\rm Re}(\lambda_k)$ (using same style code as in Fig.~\ref{ratesd}),
of the modes $\hat\rho_k$,
as functions of the distance  $d$ between the spins, for different accelerations $a$. 
 Note that for any $a>0$, the relaxation rate of mode $\hat \rho_3$ goes to zero as $d\to 0$.}
\label{ratesa}
\end{figure*}
In Fig.~\ref{ratesa}, we plot the same decoherence/relaxation rates, now as functions of the spins' separation
$d$, for different values of proper acceleration $a$.

From Eq.~(\ref{evolfinalasymp}), we see that the accelerated spin 
system eventually evolves to an equilibrium state $\hat \rho_{\rm eq} = \hat \rho_1$ given 
by Eq.~(\ref{rho1}). 
Using Eqs.~(\ref{GFstilde}-\ref{calF}), this equilibrium state,
written in terms of the elements of the Bell basis
$\{|\Psi_{\rm AB}^{(\pm)}\rangle,|\Phi_{\rm AB}^{(\pm)}\rangle\}$, reads
\begin{eqnarray}
\hat\rho_{\rm eq}& =& \frac{1}{{\cal Z}}\left[{\rm e}^{\pi J/(2a)} \left(|\Psi_{\rm AB}^{(+)}\rangle \langle\Psi_{\rm AB}^{(+)}|+|\Psi_{\rm AB}^{(-)}\rangle \langle\Psi_{\rm AB}^{(-)}|\right)
+{\rm e}^{-\pi J/(2a)} \left(|\Phi_{\rm AB}^{(+)}\rangle \langle\Phi_{\rm AB}^{(+)}|+|\Phi_{\rm AB}^{(-)}\rangle \langle\Phi_{\rm AB}^{(-)}|\right)\right]\nonumber \\
&=&\frac{{\rm e}^{-\beta\hat H_0}}{\cal Z},
\label{rhoeqa}
\end{eqnarray}
where ${\cal Z}:= 2 \left[{\rm e}^{\pi J/(2a)}+{\rm e}^{-\pi J/(2a)}\right] = {\rm tr}_{\rm s}({\rm e}^{-\beta \hat{H}_0})$ and $\beta := 
2\pi/a$.
In other words, 
for any
initial state $\hat \rho_{\rm s0}$, the final
equilibrium state $\hat \rho_{\rm eq}$ of the spin system is the 
Gibbs thermal state with  temperature $\beta^{-1} = a/(2\pi)$, the Unruh temperature.
(Observe that $|\Psi_{\rm AB}^{(+)}\rangle \langle\Psi_{\rm AB}^{(+)}|+|\Psi_{\rm AB}^{(-)}\rangle \langle\Psi_{\rm AB}^{(-)}|$ and $|\Phi_{\rm AB}^{(+)}\rangle \langle\Phi_{\rm AB}^{(+)}|+|\Phi_{\rm AB}^{(-)}\rangle \langle\Phi_{\rm AB}^{(-)}|$ are the corresponding identity matrices for the energy eigenvalue subspaces $-J/4$ and $J/4$, meaning that, as expected, the thermalization process does not favor any of the possible eigenstates of those subspaces.)
The spin system  thermalizes due to 
the vacuum fluctuations it experiences in its accelerated frame --- despite these fluctuations being distinct from the ones in inertial thermal states, as properly noted
in Ref.\cite{AS} ---, vindicating the Unruh effect also for an extended 
system.

\subsection{Spins with different proper accelerations.}
\label{subsec:diff}

\begin{figure*}
 \includegraphics[width=\textwidth,height=11cm]{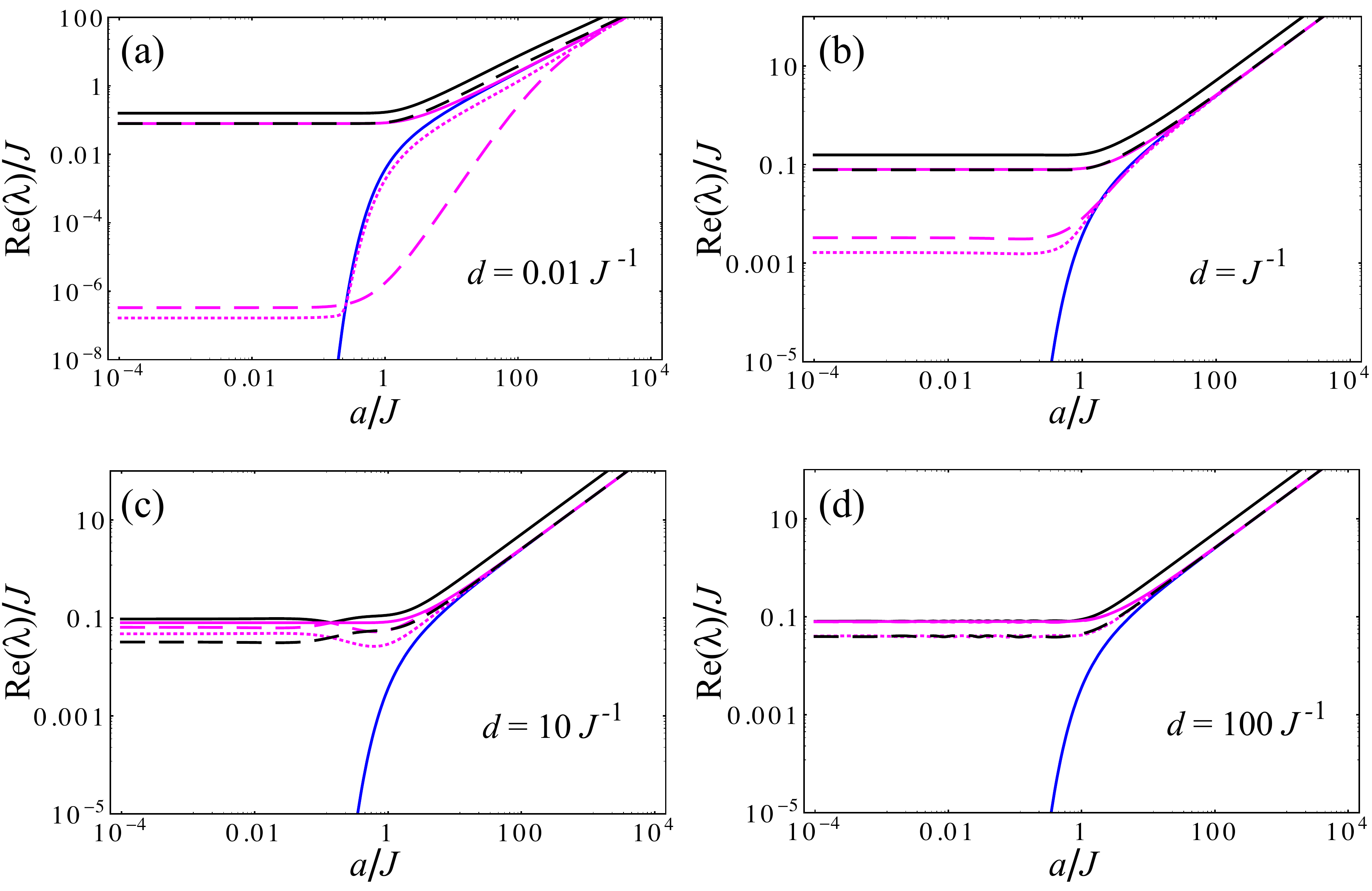}
\caption{
Acceleration dependence of relaxation/decoherence rates of the spin system with different proper accelerations.
We plot the decay rates (normalized by $q^2$ and measured by observers at spin-A location),  
${\rm Re}(\lambda_k)$, of the decaying modes of the spin system,
as functions of the spin-A acceleration $a$, for different separations $d$. We use the same style code as in Fig.~\ref{ratesd}, to which it is almost identical.}
\label{ratesdd}
\end{figure*}
\begin{figure*}
 \includegraphics[width=\textwidth,height=11cm]{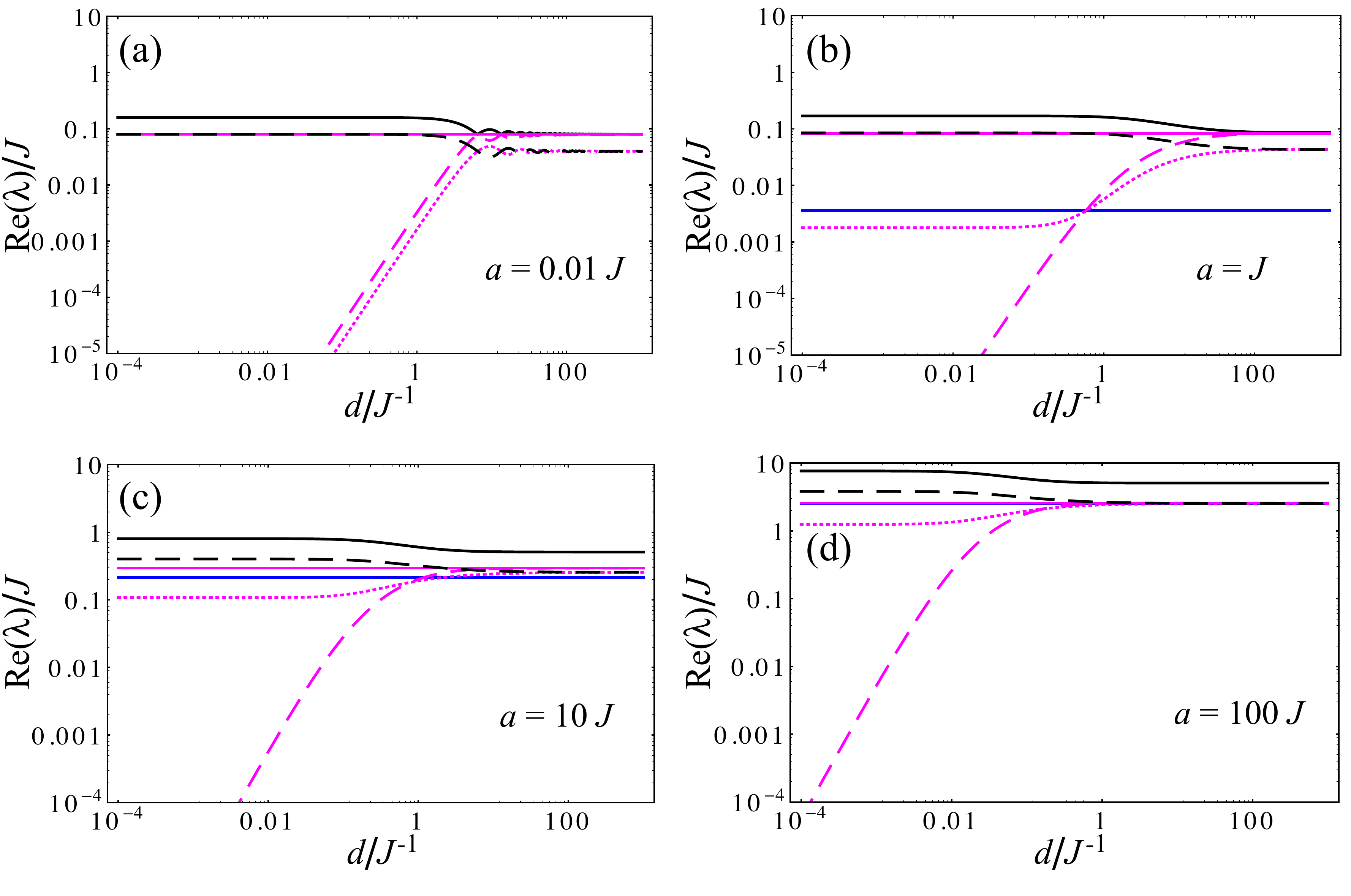}
\caption{
Spin-separation dependence of relaxation/decoherence rates of the spin system with different proper accelerations.
We plot the decay rates (normalized by $q^2$ and measured by observers at spin-A location),  ${\rm Re}(\lambda_k)$ (using same style code as in Fig.~\ref{ratesd}), of the modes $\hat\rho_k$,
as functions of the distance  $d$ between the spins, for different spin-A accelerations $a$. 
Again, note the extreme similarity with Fig.~\ref{ratesa}.}
\label{ratesad}
\end{figure*}

Now, we consider the spatial separation $d$ of the two spins to be along the direction of their accelerations:
$X_{\rm A}(\tau) \equiv Y_{\rm A}(\tau)\equiv \zeta_{\rm A}(\tau)\equiv X_{\rm B}(\tau)  \equiv Y_{\rm B}(\tau) \equiv 0$, $\zeta_{\rm B} (\tau)\equiv d$. 
In this case, $\tau \equiv \tau_{\rm A} \equiv \tau_{\rm B}/(1+ad)$  (i.e., $u_{\rm A}^0 = 1$ and $ u_{\rm B}^0 = 1/(1+ad)$) and $a$ 
continues to be the proper acceleration 
of spin A, while spin-B proper acceleration is given by
$a_{\rm B} = a/(1+ad)$ [recall discussion on uniformly-accelerated frames after
Eq.~(\ref{Rle})]. Therefore, according to the Unruh effect, each spin sees a different local temperature
at its position.

Following the same steps of the previous, equal-acceleration case, we determine
$\lambda_k$ and $\hat\rho_k$ which govern the long-term evolution of the spin system for different proper accelerations
--- see Methods' subsection ``Transformed Feynman correlators and their Principal Values.''
It turns out that the overall behavior
in this case  is very similar to the case of equal 
accelerations.
This can be readily seen from Figs.~\ref{ratesdd} and \ref{ratesad} --- where we plot the relaxation/decoherence 
rates (as measured by observers
static at $\zeta = 0$ and normalized by $q^2$) of the spin system as functions
of the (spin-A) proper acceleration $a$ and as functions of the separation $d$, respectively ---, 
which should be compared with Figs.~\ref{ratesd} and \ref{ratesa}. Note that the corresponding figures are almost identical, making  the previous discussion  on the behavior of the relaxation/decoherence rates
for $a_{\rm A}=a_{\rm B}$
also  valid for this scenario  where $a_{\rm A}\neq a_{\rm B}$. In order to better visualize the effect
of the unequal proper accelerations on the spin system, we plot in Fig.~\ref{FigDiff}
the  ratio between the relaxation/decoherence rates for different accelerations,
${\rm Re}(\lambda_\text{\rm diff})$, and for equal accelerations, ${\rm Re}(\lambda_\text{\rm eq})$, for  each decaying mode of the reduced density matrix.
We see that intermediate values of $ad$ lead to maximum differences between these two scenarios.
\begin{figure*}
 \includegraphics[width=\columnwidth,height=21cm]{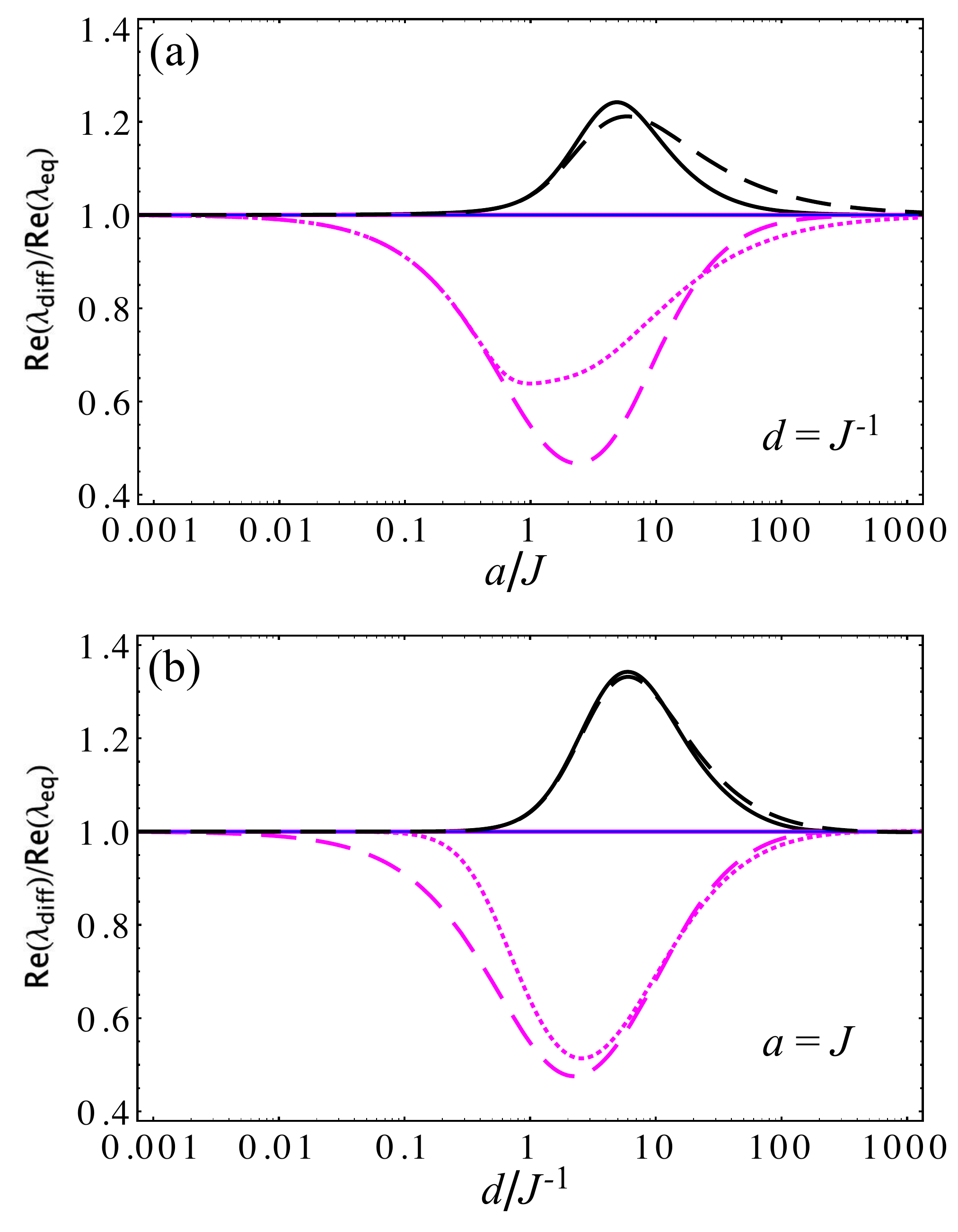}
\caption{Ratio between the relaxation/decoherence rates for different scenarios.
For each decaying mode $\hat\rho_k$ of the spin system, we 
plot (using the same style code
as in Fig.~\ref{ratesd})  the ratios ${\rm Re}(\lambda_\text{\rm diff})/{\rm Re}(\lambda_\text{\rm eq})$, where
${\rm Re}(\lambda_\text{\rm diff})$ is the decay rate in the case of spins with 
different proper accelerations and ${\rm Re}(\lambda_\text{\rm eq})$ is the decay rate in the case of spins with 
equal proper accelerations. In   (a)~the ratios are plotted as functions of spin-A's proper acceleration $a$ 
(for fixed $d = J^{-1}$), while in (b)~the rations are plotted as functions of the spins' separation $d$ (for fixed $a=J$).}
\label{FigDiff}
\end{figure*}

Most important for our purposes, though, is the fact that,
as in the case with $a_{\rm A}=a_{\rm B}$, 
the final equilibrium state $\hat\rho_1$ takes the form of  the Gibbs state given by Eq.~(\ref{rhoeqa});
i.e., the spin system thermalizes
at a temperature $\beta^{-1} = a/(2\pi)$ according to observers at $\zeta = 0$ --- for whom $\hat{H}_0$ is the free Hamiltonian of the extended system.
As for observers at $\zeta \neq 0$ (e.g.,  with spin B, $\zeta = d$) --- for whom the proper time is  $\tilde{\tau} = (1+a\zeta) \tau$ ---, 
the  same Gibbs state describes thermal equilibrium at temperature $T_{\rm U}({\zeta}) = 
a/[2 \pi(1+a\zeta)]=a_{\zeta}/(2\pi)$,
since, for them, the Hamiltonian of the spin system [i.e., the time-evolution operator appearing in Eq.~(\ref{drhodt}) with $\tau$ substituted by
$\tilde\tau/(1+a\zeta)$] is given by ${\hat H}_0/(1+a\zeta)$.
Although the observed local temperatures are different --- ensuring no net heat flow across the system, as pointed out earlier ---, the same final state 
given by Eq.~(\ref{rhoeqa}) is a true thermal 
equilibrium state for  all observers simultaneously, 
\begin{eqnarray} 
\hat{\rho}_{\rm eq} \propto {\rm exp}\left(-\frac{2\pi }{a}\hat{H}_0\right) = {\rm exp}\left(-\frac{2\pi}{a_\zeta}
\frac{ \hat{H}_0}{ (1+a\zeta)}\right),
\label{H0aH0zeta}
\end{eqnarray}
once more corroborating the physical reality of the Unruh thermal bath and its inhomogeneous temperature for an extended system.

\section{Discussion}

 We made use of an accelerated extended system (two directly coupled spins, 
each of which linearly and weakly 
 coupled to  quantum fields in the vacuum state)
 in order to show 
that the standard interpretation of the Unruh effect ---
that the inertial vacuum state acts as a legitimate thermal reservoir according to uniformly accelerated 
observers --- is strictly correct even when considering extended systems.
It is indeed true, as pointed out by Ref.\cite{AS},  that correlations 
seen by uniformly-accelerated observers in the vacuum state differ 
from the ones seen by inertial observers in an inertial thermal state.
For instance, the vacuum two-point function evaluated along two uniformly-accelerated worldlines,
in the simpler case where they have the same proper acceleration $a$ (perpendicular to their separation $d$), reads [see Eqs.~(\ref{W},\ref{sAB})]
\begin{eqnarray}
W^\bot_{{\rm ac}}(\tau,d) = -\frac{1}{4\pi^2}\frac{a^2}{\left\{4\left[\sinh\left(a\tau/2-{\rm i}\epsilon\right)\right]^2-a^2d^2\right\}},
\label{Waccelperp}
\end{eqnarray}
whereas the two-point function evaluated along two inertial wordlines 
at rest in an inertial thermal state with temperature $T$ is given by\cite{Weldon}
\begin{eqnarray}
W^{(T)}_{\rm in}(\tau,d)& =& \frac{T}{8\pi d}\left[\coth\left(\pi T (\tau+d-{\rm i}\epsilon)\right)-\coth\left(\pi T (\tau-d-{\rm i}\epsilon)\right)\right],
\label{Wthermal}
\end{eqnarray}
with $\tau$ and $d$ being, in both expressions, the  proper-time difference 
and  proper distance, respectively, measured by observers following the corresponding
worldlines. It is not difficult to verify that there is no correspondence between $a$ and $T$ which makes these
two expressions equal (as functions of $\tau$ and $d$),  except in the limit case 
$d\to 0$, for which $W^\bot_{\rm ac}(\tau,0) \equiv W^{(T_{\rm U})}_{\rm in}
(\tau,d\to 0)$,
$T_{\rm U} = a/(2\pi)$. 
This also occurs in the case of uniformly-accelerated worldlines with different proper 
accelerations, where
the vacuum two-point function evaluated along these worldlines reads [see Eqs.~(\ref{W},\ref{sABd})]
\begin{eqnarray}
W^{\parallel}_{\rm ac}(\tau,d) = -\frac{1}{4\pi^2}\frac{a^2}{\left\{4(1+ad)
\left[\sinh\left(a\tau/2-{\rm i}\epsilon\right)\right]^2-a^2d^2\right\}}.
\label{Waccelpar}
\end{eqnarray}
Again, $W^{\parallel}_{\rm ac}(\tau,d) \nequiv W^{(T)}_{\rm in}(\tau,d)$ for any fixed relation between $T$ and $a$, 
but $W^\parallel_{\rm ac}(\tau,0) \equiv W^{(T_{\rm U})}_{\rm in}
(\tau,d\to 0)$,
$T_{\rm U} = a/(2\pi)$.
\footnote{As a curiosity, though, note that  there are ``effective distances,''
$d^\bot_{\rm eff} = (2/a) \sinh^{-1}(ad/2)$
and $d^\parallel_{\rm eff} = (2/a) \sinh^{-1}(ad/\sqrt{4+4ad})$,
 such that
 $$W^{\bot}_{\rm ac}(\tau,d) \equiv \frac{a\,d^\bot_{\rm eff}}{\sinh(a\,d^\bot_{\rm eff})}\,W^{(T_{\rm U})}_{\rm in}(\tau,d^\bot_{\rm eff})$$ 
 and
 $$W^{\parallel}_{\rm ac}(\tau,d) \equiv  
 \frac{2a\,d^\parallel_{\rm eff}}{\exp({2a\,d^\parallel_{\rm eff}})-1}\,W^{(T_{\rm U})}_{\rm in}(\tau,d^\parallel_{\rm eff}).$$}
This means that, although point-like probes cannot distinguish between 
(i) being with constant proper acceleration $a$ in the vacuum and (ii) being
at rest in an inertial thermal bath with temperature $T = T_{\rm U} $, extended probes can. 
In fact, our Fig.~\ref{FigDiff} illustrates this well: 
while spatially-extended probes at rest in an inertial thermal bath cannot exhibit any 
dependence on its spatial orientation, the decoherence/relaxation
time scales of our extended system do depend on the system's orientation w.r.t.\ its acceleration --- owning to the 
fact that $W^{\parallel}_{\rm ac}(\tau,d) \nequiv W^{\bot}_{\rm ac}(\tau,d)$. 

This kind of behavior 
has been occasionally interpreted as
a violation of the strict thermal character of the Unruh effect for extended systems and non-local observables.
However --- and this is the important point ---, this distinct behavior in situations (i) and (ii) 
is  not in conflict with the rigorous statement of the Unruh effect, which only says that
 the vacuum state is a  thermal state 
according to uniformly-accelerated observers. Thermal states at the same temperature 
need not be equal, for they depend on the
Hamiltonian describing the system; and the Hamiltonian carries a subtle but important 
dependence on the family of observers
w.r.t.\ whom the time evolution is being considered. Putting it more clearly: the 
Hamiltonian of a quantum field according
to a family of inertial observers is not the same as the one according to a family
 of uniformly-accelerated observers; therefore, no need
to lead to the same expected values of similarly-defined observables. On the other 
hand, we do expect, on the grounds of the zeroth law of
thermodynamics, that any probe in contact with different thermal states at the
 same temperature reaches the same final equilibrium state
--- thermal equilibrium is  transitive. This
is exactly what happens to our accelerated extended system in the vacuum. 
It evolves towards 
the Gibbs state given by Eq.~(\ref{rhoeqa}), which is the same equilibrium state it would have
reached if it were at rest in an inertial thermal bath at temperature $T = T_{\rm U}$. This is all one can ask of
a legitimate thermal reservoir.

It is worth stressing that the model chosen here, although simple, does 
constitute a genuine extended system, since it cannot be mapped onto a two free-particle system, 
and hence non-local correlations shall be present in the dynamics. As for the equilibrium state, even though the free-system Hamiltonian presents only two eigenvalues, 
the predictions extracted from it cannot emerge from a simple two-level analysis, because the corresponding eigenenergy subspaces are spanned by states with different 
local properties. Such a character leads the expected values of local observables, e.g., $\langle\hat{s}^j_M\rangle={\rm Tr}(\hat{\rho}_{\rm s}\hat{s}^j_M)$, to be ill-determined in 
the two-level modeling. Our extended system captures all features used in the literature to argue against the physical reality of the Unruh temperature, with $T_{\rm U} = a/(2\pi)$ emerging
naturally from the thermalization process observed for the spin system.
Therefore, there is no reason to believe that our conclusions would not hold for more complex extended systems. In particular, a dramatic
 conjectured consequence
is the existence of a  critical acceleration 
above which an accelerated  magnet in the vacuum would be demagnetized, 
something which would be hard to anticipate if it were not for the Unruh effect. 
This complex situation is currently under investigation.
    
\section{Methods}

\subsection{Validity of the Markovian regime.}

The strategy of breaking down the long-term evolution of the  (open) spin system
into a sequence 
of $N$~($\gg 1$) time lapses $\Delta \tau$, such that in each time lapse, for sufficiently small coupling~$q$, 
the spins' evolution is
well described by Eq.~(\ref{rhogenfinal}),
depends on the existence of an
appropriate time lapse $\Delta \tau$ and a sequence of field states $\left\{\hat \rho_{\Phi {k}}\right\}_{{k}=0,1,\dots,N-1}$ such that
\begin{eqnarray}
{\rm tr}_\Phi \left\{
\hat\rho(\tau_{k+1}) \right\} =
{\rm tr}_\Phi \left\{\hat U_{k+1,k} \,{\rm tr}_\Phi \left[
\hat\rho(\tau_k)
\right]\otimes \hat\rho_{\Phi k}\,
\hat U_{k+1,k}^{-1}
\right\}\!\!,
\label{Mark}
\end{eqnarray}
where
$\tau_k := \tau_0+k \Delta \tau$ and $\hat U_{l,k}:=\hat U(\tau_l,\tau_k)= {\rm e}^{-{\rm i}\hat H_0 (l-k)\Delta \tau}\hat U_{\rm I}(\tau_l,\tau_k)$
(see Fig.~\ref{Spaces}). 
\begin{figure}[h]
\includegraphics[scale=0.45]{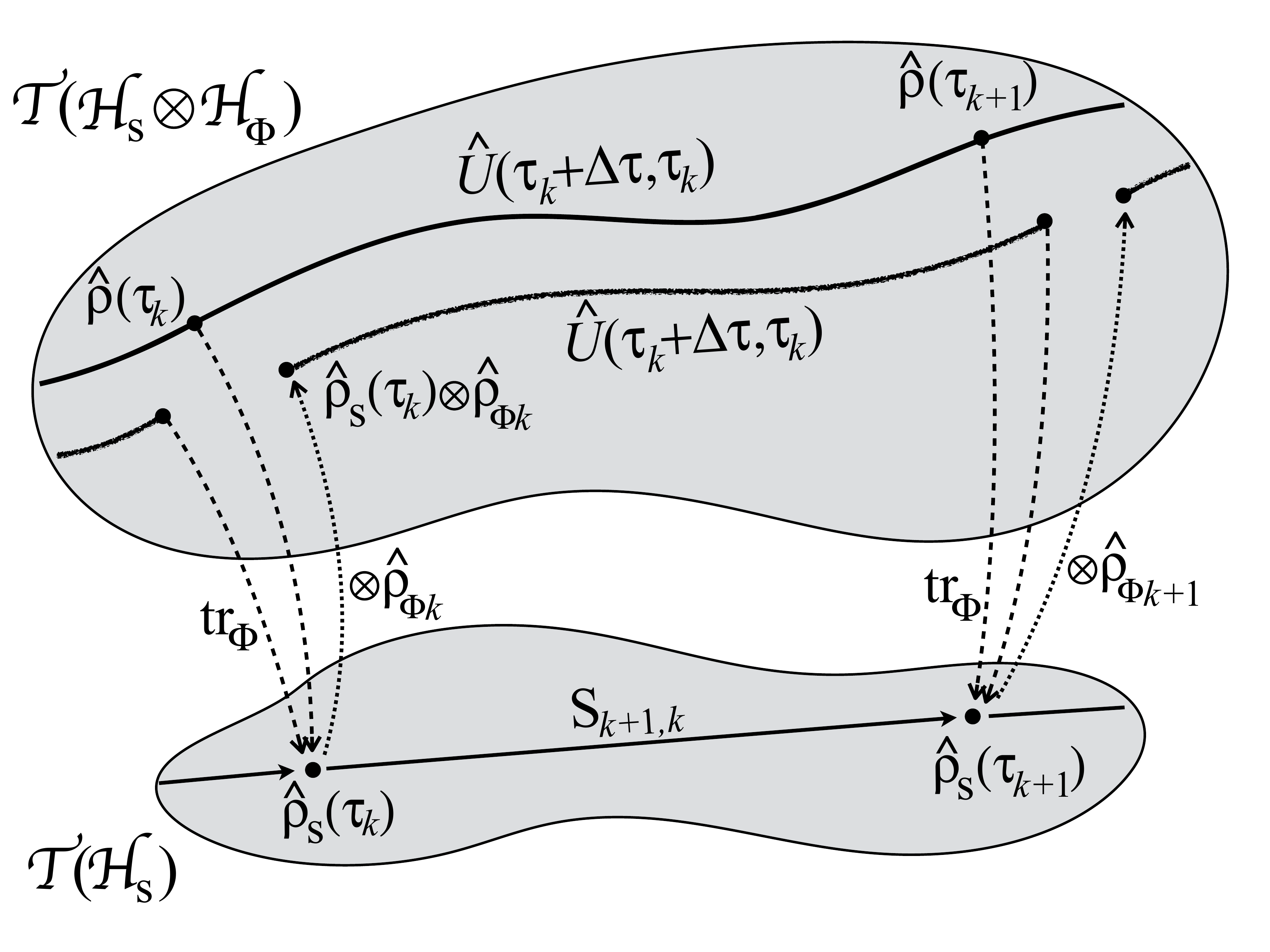}
\caption{Markovian regime and the evolution of the reduced density operator. This is a schematic representation of the
condition for the validity of the Markovian regime for the evolution of the open spin system. The full  (unitary) evolution in the space of trace-class operators describing the  universe, 
${\cal T}\left({\cal H}_{\rm s}\otimes {\cal H}_\Phi\right)$, must induce discrete dynamical maps ${\rm S}_{k+1,k}$ 
on the space of trace-class operators describing  the spin system,
${\cal T}\left({\cal H}_{\rm s}\right)$, in such a way that, for $m\geq l\geq k$,
${\rm S}_{m,l}\cdot {\rm S}_{l,k}= {\rm S}_{m,k}$ (semigroup property).}
\label{Spaces}
\end{figure}
However, the condition expressed in Eq.~(\ref{Mark}) is impracticable  since it 
assumes knowledge of the whole system evolution
$\hat \rho(\tau)$. 
Notwithstanding, we can work with a more convenient (although stronger) condition 
obtained by
defining the family of trace-preserving  maps ${\rm S}_{l,k}$,
\begin{eqnarray}
{\rm S}_{l,k}\left(\hat \rho_{\rm s}\right):=
{\rm tr}_\Phi \left\{\hat U_{l,k} \,
\hat\rho_{\rm s}\otimes \hat\rho_{\Phi k}\,
\hat U_{l,k}^{-1}
\right\},\;\;\;l\geq k,
\label{Sjk}
\end{eqnarray}
acting on the space of trace-class operators ${\cal T}\left({\cal H}_{\rm s}\right)\ni \hat \rho_{\rm s}$ describing the spin system, and asking if there is a regime (i.e., values of $\Delta \tau$ and 
$\{\hat\rho_{\Phi k}\}_{k=0,1,\dots}$) such that these maps  satisfy the composition law
${\rm S}_{m,l} \cdot{\rm S}_{l,k}= {\rm S}_{m,k}$, $m\geq l \geq k$ (semigroup property). If this can be established, then Eq.~(\ref{Mark}) 
holds for an arbitrary simply-separable initial state
$\hat\rho_0 = \hat\rho_{\rm s0}\otimes \hat \rho_{\Phi 0}$. We call this regime  Markovian, for $\hat\rho_{\rm s}(\tau_{k+1}) = {\rm S}_{k+1,k}(\hat\rho_{\rm s}(\tau_k))$.
Note, recalling   $\hat U(\tau,\tau_0) = {\rm e}^{-{\rm i}\hat H_0 (\tau-\tau_0)} \hat U_{\rm I}(\tau,\tau_0)$,  that ${\rm S}_{l,k}\left(\hat\rho_{\rm s}\right) = {\rm E}_{0}^{l-k}\left({\rm S}^{\rm I}_{l,k}\left(\hat\rho_{\rm s}\right)\right)$,
where
${\rm E}_0(\cdot):={\rm e}^{-{\rm i}\hat H_0\Delta\tau} (\cdot) {\rm e}^{{\rm i}\hat H_0\Delta\tau }$ is the free evolution on ${\cal T}\left({\cal H}_{\rm s}\right)$ and
${\rm S}^{\rm I}_{l,k}$ is given by Eq.~(\ref{Sjk}) with $\hat U$ substituted by $\hat U_{\rm I}$.

A reasonable {\it ansatz}  for the sequence $\left\{\hat \rho_{\Phi k}\right\}_{k=0,1,\dots,N-1}$ of field states is the one obtained by
applying the analogous of Eq.~(\ref{Mark}) for obtaining the reduced density matrix describing the field state; i.e., substituting, in Eq.~(\ref{Mark}),
${\rm tr}_\Phi$ by ${\rm tr}_{\rm s}$ and $\hat \rho_{\Phi k}$ by $\hat \rho_{{\rm s}k}=\hat \rho_{\rm s}(\tau_k)$. This, together with Eq.~(\ref{Mark}), 
would lead to a coupled evolution of reduced density matrices $\hat \rho_{\rm s}(\tau_k)$ and $\hat \rho_{\Phi}(\tau_k)$. In our case of interest, however,
we expect, on physical grounds, that after some transient time --- related to the time needed for the spins to exchange information via fields and the decay of the field's correlation functions  ---, the field state with which the spins interact continues to be well approximated by the initial stationary state, so that
$\hat \rho_{\Phi k} = \hat \rho_{\Phi 0}$ may provide a good candidate sequence. Indeed, the description put forward here is the one associated with the Markov approximation assumed in the context of open quantum 
systems\cite{weiss, breuer}. There, it is well established that such an approximation furnishes a good description for the system's reduced dynamics as long as the key elements are satisfied, namely, (i) the environment role is played by a large system (huge number of degrees of freedom) in a thermal state; (ii) the system-environment coupling can be considered weak; (iii)  the environment-correlation-functions time decay must be much shorter than the system evolution time scale. 

As a consequence of this approximation, 
${\rm S}_{l,k}$  only depends on $l-k$ and
the composition rule then demands
\begin{eqnarray}
 {\rm E}_0^{l-k} \cdot{\rm S}^{\rm I}_{l,k} =  \left({\rm E}_0\cdot {\rm S}_{\rm I}\right)^{l-k},
 \label{commut}
 \end{eqnarray}
where, for sufficiently small coupling $q$, $ {\rm S}_{\rm I} (\hat\rho_{\rm s}) :=  {\rm S}^{\rm I}_{k+1,k} (\hat\rho_{\rm s})\approx\left({1}
 -q^2{\rm R}_{\Delta\tau} \right)\hat\rho_{\rm s}$ can be read from the right-hand side of Eq.~(\ref{rhogenfinal}) --- substituting 
 $\hat \rho_{\rm s0}$ by
 $\hat \rho_{\rm s}$ in Eqs.~(\ref{Cpm}) and (\ref{Dpm}); the linear transformation ${\rm R}_{\Delta \tau}$, acting on ${\cal T}\left({\cal H}_{\rm s}\right)$, stands for the second-order term of Eq.~(\ref{rhogenfinal})
 for a given time lapse $\Delta \tau$.
 
Summing up, the strategy of breaking down long-term evolution of the (open) spin system  into $N$ limited time steps
 $\Delta \tau$,  for each of which Eq.~(\ref{rhogenfinal}) can be applied, depends on the validity of Eq.~(\ref{commut}) for some $\Delta \tau$.
 In particular, for $n$ time steps such that Eq.~(\ref{rhogenfinal}) can still be used for the time lapse $n\Delta \tau$,
 Eq.~(\ref{commut}) implies
 \begin{eqnarray}
{\rm R}_{n\Delta\tau} = {\rm R}_{\Delta\tau}+ {\rm E}_0^{-1}\cdot{\rm R}_{(n-1)\Delta\tau}\cdot{\rm E}_0.
 \label{commutinf}
 \end{eqnarray}

The linear transformations ${\rm E}_0$ and 
 ${\rm R}_{\Delta\tau}$  acting on the space of density matrices can be explicitly represented as
  $16 \times 16$ matrices once a basis for the spin states  and an ordering of indices
of $\hat \rho_{\rm s}$ are chosen. For instance, one could use the product states
$|\pm \rangle_{\rm A} |\pm\rangle_{\rm B}$ as elements of the basis --- where
$\hat s_M^{\rm z} |\pm\rangle_M = \pm (1/2) |\pm \rangle_M$ ---, define 
the density-matrix 
elements 
$\rho^{\alpha \beta}_{\alpha' \beta'} := \,_{\rm B}\langle\beta |\, _{\rm A}\langle \alpha| \hat \rho|\alpha' \rangle_{\rm A}  |\beta'\rangle_{\rm B}$,
and then sort these elements in a column matrix as
\begin{eqnarray}
\hat \rho = \left(
\rho^{++}_{++} \hskip .2cm \rho^{++}_{+-} \hskip .2cm\rho^{++}_{-+}  \hskip .2cm \rho^{++}_{--}  \hskip .2cm \rho^{+-}_{++}  \hskip .2cm \dots  \hskip .2cm \rho^{--}_{--}
\right)^\intercal.
\label{rho16}
\end{eqnarray}
This would lead to a particular representation of  ${\rm E}_0$ and ${\rm R}_{\Delta \tau}$.  The physical conclusions are, of course, independent of the
representation that is chosen.

It turns out that ${\rm E}_0$ and  ${\rm R}_{\Delta\tau}$ assume simpler forms when density matrices are 
expressed in the  Bell basis, formed by the
elements
\begin{eqnarray}
|\Psi_{\rm AB}^{(\pm)}\rangle &:=& \frac{1}{\sqrt{2}}\left(|+\rangle_{\rm A}|+\rangle_{\rm B}\pm |-\rangle_{\rm A}|-\rangle_{\rm B} \right),
\label{Bellbasis1}
\\
|\Phi_{\rm AB}^{(\pm)}\rangle &:=& \frac{1}{\sqrt{2}}\left(|+\rangle_{\rm A}|-\rangle_{\rm B}\pm |-\rangle_{\rm A}|+\rangle_{\rm B} \right).
\label{Bellbasis2}
\end{eqnarray}
In this representation, ${\rm E}_0$ is diagonal --- as in the product-state basis --- and ${\rm R}_{\Delta \tau}$ is ``almost diagonal'': all but 32  (out of the
240) off-diagonal terms vanish. 
In addition, 
one can explicitly check that 
the part of ${\rm R}_{\Delta\tau}$ associated with $\hat C^{(+)}_{(MN)}$ and $\hat D^{(-)}_{(MN)}$
 commutes with the free evolution ${\rm E}_0$, which is not the case for the part associated with $\hat C^{(-)}_{(MN)}$ and $\hat D^{(+)}_{(MN)}$.

Considering that the Feynman correlators appearing in Eq.~(\ref{rhogenfinal}) decrease fast enough for
$\xi \gg d$, one can verify that for a (limited) $\Delta \tau \gg J^{-1},d$, Eq.~(\ref{rhogenfinal}) assumes the asymptotic form
\begin{eqnarray}
{\rm e}^{{\rm i}\hat H_0 \Delta \tau}\hat \rho_{\rm s}(\tau)\,{\rm e}^{-{\rm i}\hat H_0 \Delta \tau}  &
\widesim[2]{\Delta\tau\gg J^{-1},d}
& 
 \hat \rho_{\rm s0}
-\frac{q^2}{2}\Delta \tau
\sum_{
{M,N\in\{{\rm A},{\rm B}\}}}
 \left\{ {\rm i} \pi\widetilde{G}_{MN}(J/2)\, \hat C^{(+)}_{(MN)} \right.
 \nonumber \\
 & &\left.-{\cal P}_{J/2}\left[{\rm i}\widetilde{G}_{MN}\right]\hat D^{(-)}_{(MN)} 
\right\}
+{\rm H.c.}
 + {\cal O} (q^3),
 \label{rhotilde}
\end{eqnarray}
where $\widetilde{G}_{MN}(J/2)$ and ${\cal P}_{J/2}[{\rm i}\widetilde{G}_{MN}]$ are defined through Eqs.~(\ref{Gftilde}) and (\ref{PV}).
Therefore, in this regime, ${\rm R}_{\Delta\tau}$ is linear in $\Delta \tau$, ${\rm R}_{\Delta\tau} \equiv  \Delta \tau\, {\rm R}_0$
--- see Eq.~(\ref{R0}) ---, and
commutes with the free evolution ${\rm E}_0$ (see remarks above). This is enough to guarantee that Eq.~(\ref{commutinf}) is satisfied and, thus, 
establish that for sufficiently small $q$, the long-term evolution of the reduced spin system is given by
\begin{eqnarray}
\hat\rho_{\rm s}(\tau_N) = {\rm S}_{N,0}\left(\hat\rho_{\rm s0}\right)
= {\rm e}^{-{\rm i} \hat H_0 N \Delta\tau}\left({\rm e}^{-q^2{\rm R}_0 N\Delta \tau}\hat\rho_{\rm s0}\right){\rm e}^{{\rm i} \hat H_0 N\Delta\tau}.
\label{evolfinal}
\end{eqnarray}

\subsection{Decaying modes and related decay rates of the spin system.
}
\label{app:eig}

Here, we present the eigenvalues $\lambda_k$ and eigenmatrices $\hat\rho_k$ of the linear operator
${\rm R}_0$, appearing in the long-term evolution above, in terms of  ${\rm i}\widetilde{G}_{MN}(J/2)$, ${\cal P}_{J/2}[{\rm i}\widetilde{G}_{MN}]$, and  the Bell states
defined in Eqs.~(\ref{Bellbasis1}) and (\ref{Bellbasis2}). First, we define the following real quantities:
\begin{eqnarray}
\alpha^\pm_{\rm s}&:=&\frac{\pi}{2}\left[{\rm Re}\left\{{\rm i}\widetilde{G}_{\rm AA}(J/2) \right\}+
{\rm Re}\left\{{\rm i}\widetilde{G}_{\rm BB}(J/2) \right\}\right]\nonumber \\
& &\pm \frac{1}{2}\left[{\rm Im}\left\{{\cal P}_{J/2}\left[{\rm i}\widetilde{G}_{\rm AA}\right]\right\}+
{\rm Im}\left\{{\cal P}_{J/2}\left[{\rm i}\widetilde{G}_{\rm BB}\right]\right\}\right],
\label{alphas}\\
\Delta\alpha^\pm_{\rm s}&:=&\frac{\pi}{2}\left[{\rm Re}\left\{{\rm i}\widetilde{G}_{\rm AA}(J/2) \right\}-
{\rm Re}\left\{{\rm i}\widetilde{G}_{\rm BB}(J/2) \right\}\right]\nonumber \\
& &\pm \frac{1}{2}\left[{\rm Im}\left\{{\cal P}_{J/2}\left[{\rm i}\widetilde{G}_{\rm AA}\right]\right\}-
{\rm Im}\left\{{\cal P}_{J/2}\left[{\rm i}\widetilde{G}_{\rm BB}\right]\right\}\right],
\label{dalphas}
\end{eqnarray}
\begin{eqnarray}
\alpha^\pm_{\rm i}&:=&\pi{\rm Re}\left\{{\rm i}\widetilde{G}_{\rm AB}(J/2) \right\}
\pm{\rm Im}\left\{{\cal P}_{J/2}\left[{\rm i}\widetilde{G}_{\rm AB}\right]\right\},
\label{alphai}
\\
\beta^\pm_{\rm s}&:=&\frac{\pi}{2}\left[{\rm Im}\left\{{\rm i}\widetilde{G}_{\rm AA}(J/2) \right\}+
{\rm Im}\left\{{\rm i}\widetilde{G}_{\rm BB}(J/2) \right\}\right]\nonumber \\
& &\pm \frac{1}{2}\left[{\rm Re}\left\{{\cal P}_{J/2}\left[{\rm i}\widetilde{G}_{\rm AA}\right]\right\}+
{\rm Re}\left\{{\cal P}_{J/2}\left[{\rm i}\widetilde{G}_{\rm BB}\right]\right\}\right],
\label{betas}\\
\Delta\beta^\pm_{\rm s}&:=&\frac{\pi}{2}\left[{\rm Im}\left\{{\rm i}\widetilde{G}_{\rm AA}(J/2) \right\}-
{\rm Im}\left\{{\rm i}\widetilde{G}_{\rm BB}(J/2) \right\}\right]\nonumber \\
& &\pm \frac{1}{2}\left[{\rm Re}\left\{{\cal P}_{J/2}\left[{\rm i}\widetilde{G}_{\rm AA}\right]\right\}-
{\rm Re}\left\{{\cal P}_{J/2}\left[{\rm i}\widetilde{G}_{\rm BB}\right]\right\}\right],
\label{dbetas}\\
\beta^\pm_{\rm i}&:=&\pi{\rm Im}\left\{{\rm i}\widetilde{G}_{\rm AB}(J/2) \right\}\pm{\rm Re}\left\{{\cal P}_{J/2}\left[{\rm i}\widetilde{G}_{\rm AB}\right]\right\},
\label{betai}
\end{eqnarray}
from which all ${\rm i}\widetilde{G}_{MN}(J/2)$ and ${\cal P}_{J/2}\left[{\rm i}\widetilde{G}_{MN}\right]$ 
can be reconstructed.
Also, let $r_0$, $r_\pm$  be the roots of the polynomial
\begin{eqnarray}
P(r)&:=& 
r^3-2( \alpha^-_{\rm s}+2 \alpha^+_{\rm s}) r^2+4\left[(\alpha^+_{\rm s})^2+2\alpha^-_{\rm s} \alpha^+_{\rm s}+ (\beta^-_{\rm i})^2
-\Delta\alpha^-_{\rm s} \Delta\alpha^+_{\rm s}\right] r\nonumber \\
& & 
-8 \alpha^-_{\rm s} \left[(\alpha^+_{\rm s})^2+ (\beta^-_{\rm i})^2\right]+
8 \alpha^+_{\rm s} \Delta\alpha^-_{\rm s} \Delta\alpha^+_{\rm s},
\label{pr}
\end{eqnarray}
ordered, by convention, in such a way that they are continuous functions
of $\Delta\alpha^+_{\rm s}\Delta\alpha^-_{\rm s}$ and, for small enough
$\Delta\alpha^+_{\rm s}\Delta\alpha^-_{\rm s}$, $r_0\approx 2\alpha^-_{\rm s}+{\cal O}(\Delta\alpha^+_{\rm s}\Delta\alpha^-_{\rm s})$, $r_\pm
\approx 2(\alpha^+_{\rm s}\pm i\beta^-_{\rm i})+{\cal O}(\Delta\alpha^+_{\rm s}\Delta\alpha^-_{\rm s})$.
Then, the eigenvalues and eigenmatrices are:
\begin{eqnarray}
\lambda_1 &=& 0,\label{l1}
\\
\hat \rho_1 &=&\frac{1}
{2\left[(\alpha_{\rm s}^+)^2-(\alpha_{\rm i}^+)^2+\alpha_{\rm s}^+\alpha_{\rm s}^- -\alpha_{\rm i}^+\alpha_{\rm i}^- \right]}
\left\{\left[(\alpha_{\rm s}^+)^2-(\alpha_{\rm i}^+)^2\right] \left(
|\Psi_{\rm AB}^{(+)}\rangle \langle\Psi_{\rm AB}^{(+)}|+|\Psi_{\rm AB}^{(-)}\rangle \langle\Psi_{\rm AB}^{(-)}|\right)\right.
\nonumber \\
& &
\left.
+ 
(\alpha_{\rm s}^++\alpha_{\rm i}^+)(\alpha_{\rm s}^--\alpha_{\rm i}^-)\,
|\Phi_{\rm AB}^{(-)}\rangle \langle\Phi_{\rm AB}^{(-)}|
+ 
(\alpha_{\rm s}^+-\alpha_{\rm i}^+)(\alpha_{\rm s}^-+\alpha_{\rm i}^-) \,
|\Phi_{\rm AB}^{(+)}\rangle \langle\Phi_{\rm AB}^{(+)}|
\right\};\,\,\,\,\,
\label{rho1}
\\
\lambda_2 &=& 2\alpha_{\rm s}^++\alpha^-_{\rm s}+ \sqrt{(\alpha^-_{\rm s})^2+4 \alpha^+_{\rm i}(\alpha^+_{\rm i}+\alpha^-_{\rm i})},\label{l2}
\\
\hat \rho_2 &=&
|\Psi_{\rm AB}^{(+)}\rangle \langle\Psi_{\rm AB}^{(+)}|+|\Psi_{\rm AB}^{(-)}\rangle \langle\Psi_{\rm AB}^{(-)}|
-\left[1+\frac{\sqrt{(\alpha^-_{\rm s})^2+4 \alpha^+_{\rm i}(\alpha_{\rm i}^+ + \alpha_{\rm i}^-)}-\alpha_{\rm s}^-}{2\alpha^+_{\rm i}}\right]
|\Phi_{\rm AB}^{(+)}\rangle \langle\Phi_{\rm AB}^{(+)}|\nonumber \\
& &\hskip 4.2cm
-\left[1-\frac{\sqrt{(\alpha^-_{\rm s})^2+4 \alpha^+_{\rm i}(\alpha_{\rm i}^+ + \alpha_{\rm i}^-)}-\alpha_{\rm s}^-}{2\alpha^+_{\rm i}}\right]
|\Phi_{\rm AB}^{(-)}\rangle \langle\Phi_{\rm AB}^{(-)}|;
\label{rho2}\\
\lambda_3 &=& 2\alpha_{\rm s}^++\alpha^-_{\rm s}- \sqrt{(\alpha^-_{\rm s})^2+4 \alpha^+_{\rm i}(\alpha^+_{\rm i}+\alpha^-_{\rm i})},\label{l3}
\\
\hat \rho_3 &=&
|\Psi_{\rm AB}^{(+)}\rangle \langle\Psi_{\rm AB}^{(+)}|+|\Psi_{\rm AB}^{(-)}\rangle \langle\Psi_{\rm AB}^{(-)}|
-\left[1-\frac{\sqrt{(\alpha^-_{\rm s})^2+4 \alpha^+_{\rm i}(\alpha_{\rm i}^+ + \alpha_{\rm i}^-)}+\alpha_{\rm s}^-}{2\alpha^+_{\rm i}}\right]
|\Phi_{\rm AB}^{(+)}\rangle \langle\Phi_{\rm AB}^{(+)}|\nonumber \\
& &\hskip 4.2cm
-\left[1+\frac{\sqrt{(\alpha^-_{\rm s})^2+4 \alpha^+_{\rm i}(\alpha_{\rm i}^+ + \alpha_{\rm i}^-)}+\alpha_{\rm s}^-}{2\alpha^+_{\rm i}}\right]
|\Phi_{\rm AB}^{(-)}\rangle \langle\Phi_{\rm AB}^{(-)}|;
\label{rho3}\\
\lambda_4 &=& 2\alpha_{\rm s}^-=\lambda_5,
\label{l45}
\\
\hat \rho_4 &=&|\Psi_{\rm AB}^{(+)}\rangle \langle\Psi_{\rm AB}^{(+)}|
-|\Psi_{\rm AB}^{(-)}\rangle \langle\Psi_{\rm AB}^{(-)}|,
\label{rho4}
\\
\hat\rho_5&=&{\rm i}|\Psi_{\rm AB}^{(+)}\rangle \langle\Psi_{\rm AB}^{(-)}|-{\rm i}|\Psi_{\rm AB}^{(-)}\rangle \langle\Psi_{\rm AB}^{(+)}|;
\label{rho5}\\
\lambda_{6}&=&r_0 ,
\label{l6}
\\
\hat\rho_{6}&=&|\Psi_{\rm AB}^{(+)}\rangle \langle\Psi_{\rm AB}^{(-)}|+|\Psi_{\rm AB}^{(-)}\rangle \langle\Psi_{\rm AB}^{(+)}|+2\Delta\alpha^-_{\rm s}\left[\frac{
|\Phi_{\rm AB}^{(+)}\rangle \langle\Phi_{\rm AB}^{(-)}|}{r_0-2(\alpha^+_{\rm s}+{\rm i}\beta^-_{\rm i})}+\frac{|\Phi_{\rm AB}^{(-)}\rangle \langle\Phi_{\rm AB}^{(+)}|}{r_0-2(\alpha^+_{\rm s}-{\rm i}\beta^-_{\rm i})}
\right];
\label{rho6}
\\
\lambda_7 &=& \alpha^+_{\rm s}+\alpha^-_{\rm s}-\alpha^+_{\rm i}+{\rm i}(\beta^+_{\rm s}-\beta^-_{\rm s}+\beta^-_{\rm i})=\lambda_8^\ast
=\lambda_9 = \lambda_{10}^\ast,
\label{l78910}
\\
\hat \rho_7 &=&|\Psi_{\rm AB}^{(-)}\rangle \langle\Phi_{\rm AB}^{(-)}|=\hat\rho_8^\dagger,
\label{rho78}
\\
\hat \rho_{9} &=&|\Psi_{\rm AB}^{(+)}\rangle \langle\Phi_{\rm AB}^{(-)}|=\hat\rho_{10}^\dagger;
\label{rho910}
\\
\lambda_{11} &=& \alpha^+_{\rm s}+\alpha^-_{\rm s}+\alpha^+_{\rm i}+{\rm i}(\beta^+_{\rm s}-\beta^-_{\rm s}-\beta^-_{\rm i})=\lambda_{12}^\ast
=\lambda_{13}=\lambda_{14}^\ast,
\label{l11121314}
\\
\hat \rho_{11} &=&|\Psi_{\rm AB}^{(-)}\rangle \langle\Phi_{\rm AB}^{(+)}|=\hat\rho_{12}^\dagger,
\label{rho1112}
\\
\hat \rho_{13} &=&|\Psi_{\rm AB}^{(+)}\rangle \langle\Phi_{\rm AB}^{(+)}|=\hat\rho_{14}^\dagger;
\label{rho1314}
\\
\lambda_{15}&=& r_-,
\label{l15}
\\
\hat\rho_{15}&=&\Delta\alpha^+_{\rm s}\left[|\Psi_{\rm AB}^{(+)}\rangle \langle\Psi_{\rm AB}^{(-)}|+|\Psi_{\rm AB}^{(-)}\rangle \langle\Psi_{\rm AB}^{(+)}|\right]
+2\Delta\alpha^+_{\rm s}\Delta\alpha^-_{\rm s}\left[\frac{
|\Phi_{\rm AB}^{(+)}\rangle \langle\Phi_{\rm AB}^{(-)}|}{r_--2(\alpha^+_{\rm s}+{\rm i}\beta^-_{\rm i})}+\frac{|\Phi_{\rm AB}^{(-)}\rangle \langle\Phi_{\rm AB}^{(+)}|}{r_--2(\alpha^+_{\rm s}-{\rm i}\beta^-_{\rm i})}
\right];\nonumber \\
\label{rho15}
\\
\lambda_{16}&=& r_+,
\label{l16}
\\
\hat\rho_{16}&=&\Delta\alpha^+_{\rm s}\left[|\Psi_{\rm AB}^{(+)}\rangle \langle\Psi_{\rm AB}^{(-)}|+|\Psi_{\rm AB}^{(-)}\rangle \langle\Psi_{\rm AB}^{(+)}|\right]+2\Delta\alpha^+_{\rm s}\Delta\alpha^-_{\rm s}\left[\frac{
|\Phi_{\rm AB}^{(+)}\rangle \langle\Phi_{\rm AB}^{(-)}|}{r_+-2(\alpha^+_{\rm s}+{\rm i}\beta^-_{\rm i})}+\frac{|\Phi_{\rm AB}^{(-)}\rangle \langle\Phi_{\rm AB}^{(+)}|}{r_+-2(\alpha^+_{\rm s}-{\rm i}\beta^-_{\rm i})}
\right].\nonumber \\
\label{rho16}
\end{eqnarray}
We note that in case $\Delta \alpha_{\rm s}^{+} \Delta \alpha_{\rm s}^{-} = 0$ --- which encompasses  both scenarios analyzed in this work (equal and different spins' proper accelerations) ---, 
 Eqs.~(\ref{l6},\ref{rho6},\ref{l15},\ref{rho15},\ref{l16},\ref{rho16}) can be conveniently reduced to
$\lambda_6 = 2\alpha^-_{\rm s}$, $\hat\rho_6 = |\Psi_{\rm AB}^{(+)}\rangle \langle\Psi_{\rm AB}^{(-)}|+|\Psi_{\rm AB}^{(-)}
\rangle \langle\Psi_{\rm AB}^{(+)}|$, $\lambda_{15} = 2(\alpha^+_{\rm s}-{\rm i}\beta^-_{\rm i})= \lambda_{16}^\ast$, 
$\hat\rho_{15} = |\Phi_{\rm AB}^{(-)}\rangle \langle\Phi_{\rm AB}^{(+)}| = \hat\rho_{16}^\dagger$.

Note that the
``mode''  associated with the null eigenvalue,  $ \hat \rho_1$  --- which gives the final equilibrium state 
in the long-term evolution of the spin system; see Eg.~(\ref{evolfinalasymp}) ---, is diagonal in the Bell basis.
Therefore, regardless the form of the Feynman correlator $G_{MN}$ --- 
provided ${\rm Re}(\lambda_{k\neq1})>0$
---, the spin system evolves to a statistical mixture of Bell states, with populations which depend on the specific
form of $G_{MN}$.
Notice from Eq.~(\ref{rho1}), however, that 
$|\Psi_{\rm AB}^{(\pm)}\rangle$ are equally populated regardless the form of $G_{MN}$, which means that the equilibrium state is also 
a statistical mixture of the separable states $|+ \rangle_{\rm A} |+\rangle_{\rm B}$ and  $|- \rangle_{\rm A} |-\rangle_{\rm B}$. The same is not true for 
$|\Phi_{\rm AB}^{(\pm)}\rangle$: depending on $G_{MN}$, the final equilibrium state may preserve correlations between
$|+ \rangle_{\rm A} |-\rangle_{\rm B}$ and  $|- \rangle_{\rm A} |+\rangle_{\rm B}$. These results can be summarized as follows: 
in general, the spin system  will loose coherence in any basis which diagonalizes,
simultaneously, the free Hamiltonian $\hat H_0$ and the
 total spin $\hat {\bf S}^2 := \sum_{j\in\{{\rm x},{\rm y},{\rm z}\}}\left(\hat s_{\rm A}^j+\hat s_{\rm B}^j\right)^2$.

\subsection{Transformed Feynman correlators and their Principal Values.}

Here, we calculate the quantities ${\rm i}\widetilde{G}_{MN}$ and ${\cal P}_{a}[{\rm i}\widetilde{G}_{MN}]$ 
which
completely determine the  long-term evolution of the spin system through Eqs.~(\ref{evolfinalasymp},\ref{alphas}-\ref{rho16}).
Treating first the case of spins with the equal proper accelerations,  Eq.~(\ref{sigma2}) leads to
\begin{eqnarray}
\sigma(x_{\rm A},x_{\rm A}')= \sigma(x_{\rm B},x_{\rm B}') &=& -\frac{4}{a^2}\left[\sinh\left(\frac{a\xi}{2}\right)\right]^2,
\label{sAA}
\\
\sigma(x_{\rm A},x_{\rm B}')= \sigma(x_{\rm B},x_{\rm A}') &=& -\frac{4}{a^2}\left[\sinh\left(\frac{a\xi}{2}\right)\right]^2+d^2,\;\;
\label{sAB}
\end{eqnarray}
where $\xi = \tau- \tau'$. Substituting these expressions into Eq.~(\ref{W}) and using that
 ${\rm i}G_{MN}(\xi) = \theta(\xi) W(x_M,x'_N)+\theta(-\xi) W(x'_N,x_M)$ --- where $\theta(\xi)$ is the 
Heaviside
step function ---, we obtain, using Eqs.~(\ref{Gftilde}) and (\ref{PV}), the quantities 
\begin{eqnarray}
{\rm i}\widetilde{G}_{\rm AA}(\omega)&=&{\rm i}\widetilde{G}_{\rm BB}(\omega) =\frac{1}{8\pi^2} 
\left[
\omega\coth
\left(\frac{\pi\omega}{a}\right)-{\rm i} \lim_{\epsilon \to 0+}\frac{1}{\epsilon}\right],
\label{GFstilde}
\\
{\rm i}\widetilde{G}_{\rm AB}(\omega)& =&\frac{\sin\left(
\frac{2\omega}{a} {\rm sinh}^{-1} \left(\frac{ad}{2}\right)\right)}{4\pi^2d\,\sqrt{4+a^2 d^2}} 
\left[
\coth
\left(\frac{\pi\omega}{a}\right)
-{\rm i} \cot\left(
\frac{2\omega}{a} {\rm sinh}^{-1} \left(\frac{ad}{2}\right)\right)\right].
\label{GFitilde}
\end{eqnarray}
The complex infinite in Eq.~(\ref{GFstilde}) is a consequence of the ``too-singular'' (ultraviolet)  behavior of $G_{MM}(\xi)$ at the
vertex of the light cone. One could ``smooth'' this singularity by smearing out the position of the spins. However,
this is not necessary for our purposes since it does not affect any physical result [notice, from Eqs.~(\ref{alphas})-(\ref{betai}) that this divergence only contributes ---
equally --- to $\beta^\pm_{\rm s}$, which, by their turn, only appear in Eqs.~(\ref{l78910}) and (\ref{l11121314}), in such a way that the divergences cancel out].
Then, with the help of Eq.~(4.115.8) of Ref.\cite{Grad}, we can calculate
${\cal P}_{J/2}\left[{\rm i}{\widetilde G}_{\rm AA}\right]={\cal P}_{J/2}\left[{\rm i}{\widetilde G}_{\rm BB}\right]$ and ${\cal P}_{J/2}\left[{\rm i}{\widetilde G}_{\rm AB}\right]$  --- the former being  obtained from the limit
$d\to 0_+$ of the latter:
\begin{eqnarray}
{\cal P}_{J/2}\left[{\rm i}{\widetilde G}_{\rm AA}\right] &=&{\cal P}_{J/2}\left[{\rm i}{\widetilde G}_{\rm BB}\right]=\frac{J}{8\pi^2} 
\left(-
\lim_{\epsilon \to 0_+} \ln\epsilon +{\rm i}\frac{ \pi}{2}\right),
\label{PGFstilde}
\\
{\cal P}_{J/2}\left[{\rm i}{\widetilde G}_{\rm AB}\right]  &=&\frac{1}{4\pi^2d\,\sqrt{4+a^2 d^2}} 
\left[
{\cal F}\left(\frac{J}{2a},{\rm sinh}^{-1}\left(\frac{ad}{2}\right)\right)
+{\rm i}\pi \sin\left(
\frac{J}{a} {\rm sinh^{-1}} \left(\frac{ad}{2}\right)\right)\right],
\label{PGFitilde}
\end{eqnarray}
where
\begin{eqnarray}
{\cal F}(x,y) &:=& -\frac{1}{x}+\pi \cos\left(2xy\right) \coth \left(\pi x\right)-2x\sum_{n=1}^\infty\frac{{\rm e}^{-2ny}}{n^2+x^2}
\nonumber \\
&=&-\frac{1}{x}+\pi \cos\left(2xy\right) \coth \left(\pi x\right)
-{\rm Re}\left\{{\rm i} \,{\rm e}^{2{\rm i}xy}B_{{\rm e}^{-2y}}\left(1+{\rm i}x,0\right)\right\}
\label{calF}
\end{eqnarray}
and $B_z(x,y) := \int_0^z dt\,t^{x-1} (1-t)^{y-1}$ is the incomplete Euler $\beta$ function.
Again,
 a  divergence related to the singular behavior of  $G_{MM}(\xi)$ at $\xi = 0$ appears
in Eq.~(\ref{PGFstilde}). This time, however, it is not obvious that this divergence will bear no consequence on 
physical results. And 
in 
fact, although the main features of the spins' evolution (the final equilibrium state and  relaxation/decoherence time scales) are completely oblivious to such a divergence, some
transient observables (e.g., the frequency of oscillation of some decaying modes) do depend on the real part of 
${\cal P}_{J/2}\left[{\rm i}{\widetilde G}_{MM}\right]$ [see, again, Eqs.~(\ref{betas}), (\ref{l78910}), and (\ref{l11121314})].
The equations above completely determine
$\lambda_k$ and $\hat\rho_k$ appearing in  the long-term 
evolution of the spin system in the case of equal proper accelerations --- see Eqs.~(\ref{evolfinalasymp}) and (\ref{alphas}-\ref{rho16}).

In the case of spins with different proper accelerations,
substituting the spins' worldlines into Eq.~(\ref{sigma2}), Eqs.~(\ref{sAA}) and (\ref{sAB}) get replaced by:
\begin{eqnarray}
\sigma(x_{\rm A},x_{\rm A}')= \frac{\sigma(x_{\rm B},x_{\rm B}')}{(1+ad)^2} &=& -\frac{4}{a^2}\left[\sinh\left(\frac{a\xi}{2}\right)\right]^2,
\label{sAAd}
\\
\sigma(x_{\rm A},x_{\rm B}')= \sigma(x_{\rm B},x_{\rm A}') &=& -\frac{4(1+ad)}{a^2}\left[\sinh\left(\frac{a\xi}{2}\right)\right]^2+d^2.
\label{sABd}
\end{eqnarray}
Applying      the same procedure above to these results (recalling that  $u_{\rm A}^0 = 1$ and $u_{\rm B}^0 = 1/(1+ad)$), we obtain
\begin{eqnarray}
{\rm i}\widetilde{G}_{\rm AA}(\omega)&=&{\rm i}\widetilde{G}_{\rm BB}(\omega) 
=\frac{1}{8\pi^2} 
\left[\omega
\coth
\left(\frac{\pi \omega}{a}\right)-{\rm i} \lim_{\epsilon \to 0+}\frac{1}{\epsilon}\right],
\label{GFstildeddif}
\\ 
{\rm i}\widetilde{G}_{\rm AB}(\omega)& =&\frac{(1+ad)\sin\left(
\frac{2\omega}{a} {\rm sinh}^{-1} \left(\frac{ad}{2\sqrt{1+ad}}\right)\right)}{4\pi^2d\,(2+a d)} 
\left[
\coth
\left(\frac{\pi \omega}{a}\right)-{\rm i} \cot\left(
\frac{2\omega}{a} {\rm sinh}^{-1} \left(\frac{ad}{2\sqrt{1+ad}}\right)\right)\right],\nonumber \\
\label{GFitildeddif}\\
{\cal P}_{J/2}\left[{\rm i}{\widetilde G}_{\rm AA}\right] &=&{\cal P}_{J/2}\left[{\rm i}{\widetilde G}_{\rm BB}\right]
=\frac{J}{8\pi^2} 
\left(-
\lim_{\epsilon \to 0_+} \ln\epsilon +{\rm i}\frac{ \pi}{2}\right),
\label{PGFstildeddif}\\
{\cal P}_{J/2}\left[{\rm i}{\widetilde G}_{\rm AB}\right]  &=&\frac{(1+ad)}{4\pi^2d\,(2+a d)} 
\nonumber\\
& &\times
\left[
{\cal F}\left(\frac{J}{2a},{\rm sinh}^{-1}\left(\frac{ad}{2\sqrt{1+ad}}\right)\right)
+{\rm i}{\rm \pi} \sin\left(
\frac{J}{a} {\rm sinh^{-1}} \left(\frac{ad}{2\sqrt{1+ad}}\right)\right)\right],
\label{PGFitildeddif}
\end{eqnarray}
where ${\cal F}$ is still given by Eq.~(\ref{calF}).
Again, these quantities  determine all the eigenvalues $\lambda_k$ and eigenmatrices $\hat\rho_k$
which govern the long-term evolution of the spin system, now in the case of different proper accelerations.

\section*{Acknowledgements} C.\ L.\ acknowledges full financial  support from S\~ao Paulo  Research Foundation (FAPESP) through Grant No.\ 2012/24728-0. F.\ B.\ is 
supported by the Instituto Nacional de Ci\^encia e Tecnologia - Informa\c{c}\~ao Qu\^antica (INCT-IQ).
J.\  H.\  was supported by Conselho Nacional de Desenvolvimento Cient\'\i fico e Tecnol\'ogico (CNPq), Grants No.\
307548/2015-5 and 312352/2018-2, and FAPESP, Grants No.\ 2015/23849-7
and No.\ 2016/10826-1. D.~V.\ acknowledges partial supported from FAPESP Grant No.\ 2013/12165-4. D.~V.\  also thanks George Matsas and
Andr\'e Landulfo for discussions in the early stages of this work. The 
authors also thank the anonymous reviewers
for important comments which helped to improve this manuscript and Prof.\ Sujoi Modak for bringing 
to our attention
Refs.\cite{Espagnat,MOPS} on proper and improper mixed states.

\newpage 


\end{document}